\documentclass[review]{elsarticle}

\usepackage{lineno}
\modulolinenumbers[5]

\usepackage{amsmath,bm}
\usepackage{graphicx}
\usepackage[colorinlistoftodos]{todonotes}
\usepackage{subcaption}
\usepackage{makecell}

\usepackage{multicol}
\usepackage{mwe}
\usepackage{amsfonts}
\usepackage[normalem]{ulem} 

\newcommand{\Xin}{X_{\text{in}}}
\newcommand{\Xout}{X_{\text{out}}}

\journal{Journal of Computational Physics}

\begin{document}

\begin{frontmatter}

\title{Predicting dynamical system evolution with residual neural networks}

\author[kiam,sk]{Artem Chashchin\corref{mycorrespondingauthor}}
\cortext[mycorrespondingauthor]{Corresponding author}
\ead{artem.chashchin@skolkovotech.ru}

\author[kiam]{Mikhail Botchev}

\author[sk,inm]{Ivan Oseledets}

\author[sk]{George Ovchinnikov}



\address[kiam]{Keldysh Institute of Applied Mathematics of the Russian Academy of Sciences, Moscow, Russia}
\address[sk]{Skolkovo Institute of Science and Technology, Moscow, Russia}
\address[inm]{Marchuk Institute of Numerical Mathematics of the Russian Academy of Sciences, Moscow, Russia}

\begin{abstract}
Forecasting time series and time-dependent data is a common problem in many applications. One typical example is solving ordinary differential equation (ODE) systems $\dot{x}=F(x)$. Oftentimes the right hand side function $F(x)$ is not known explicitly and the ODE system is described by solution samples taken at some time points. Hence, ODE solvers cannot be used. In this paper, a data-driven approach to learning the evolution of dynamical systems is considered. We show how by training neural networks with ResNet-like architecture on the solution samples, models can be developed to predict the ODE system solution further in time.
By evaluating the proposed approaches on three test ODE systems, we demonstrate that the neural network models are able to reproduce the main dynamics of the systems qualitatively well.  Moreover, the predicted solution remains stable for much longer times than for other currently known models.

\end{abstract}


\end{frontmatter}


\section{Introduction}
\label{sec:intro}

Neural network techniques are becoming an important tool for analyzing time-dependent data sets and multivariate time series. One typical problem is to reconstruct solution of an ODE (ordinary differential equation) system $\dot{\bm{x}}=F(\bm{x})$ by learning the right hand side function $F(\bm{x})$ with a suitable neural network. In~\cite{trischler2016synthesis} promising results for this problem are reported with a shallow neural network applied to learn $F(\bm{x})$ and then transformed into a recurrent neural network (RNN).

In \cite{brunton2016discovering}, this task is formulated as a sparse regression problem.
$F(\bm{x})$ is approximated with a linear combination of functions from some set and, given measurements of $\bm{x}$ and (or estimates of) $\dot{\bm{x}}$,
coefficients are recovered by a sequential thresholded least squares algorithm
or the LASSO (least absolute shrinkage and selection operator) method.
The authors also note that their approach resembles Dynamic Mode Decomposition (DMD) \cite{tu2014dynamic} in case of discrete-time dynamical systems.

However, in real life applications the right hand side function $F(\bm{x})$ may not be known explicitly and the ODE system may only be described by sampled data points. In this case the approaches of~\cite{trischler2016synthesis,brunton2016discovering} are not applicable. That is where machine learning, as a means of building models from the data, might be useful.
A possible solution for this problem is to train a model that treats the data as a time series and either reconstructs the equations or predicts the evolution of the system for a certain time
$\Delta t>0$ ahead of the current time moment $t$.
In the latter case, when applied to the data recursively several (say, $k$) times, the model should be able to predict the evolution for $k$ time steps
on the time interval $[t,t+k\Delta t]$.

In both cases, while either reconstructing $F(\bm{x})$ or
predicting the ODE solution,
a question of the accuracy of the neural network approximations
arises.

In particular, if a system is chaotic, it may be difficult to obtain small errors for long-term predictions
due to the stability effects:
even initially small differences between the original and predicted data may strongly increase with time.
It is then reasonable to require from the neural network approximation that at least the main dynamics of the system is
captured properly,
so that the predicted trajectories are not too far off the true solution.

For moderate prediction times promising results for this problem
are reported in~\cite{mai2016reconstruction}.
There a numerical method is used to reconstruct the system right hand side $F(\bm{x})$.
A favorable error behavior is observed for long-term prediction
of one- and two-dimensional (1D and 2D) problems.
However, for a three-dimensional (3D) Lorenz system exhibiting chaotic dynamics, the predicted and original trajectories
start to significantly diverge starting from relatively early times
$t\approx1$.

Consider now the second option: prediction of the ODE solution
without making any assumptions on the right hand side function
$F(\bm{x})$.
A suitable approximation model can be applied in problems where solution prediction (with a good accuracy) is more important than explicitly reconstructing the equations.
The output of the model, which is a prediction $\tilde{\bm{x}}(t+\Delta t)$ to
the ODE solution $\bm{x}(t+\Delta t)$,
is obtained as  $\tilde{\bm{x}}(t+\Delta t) = g(\bm{x}(t), \bm{\theta})$, where $\bm{\theta}$ is a vector of
model
parameters and $\bm{x}(t)$ is its input (i.e., system
state at time $t$).
There are many options for architectures to define $g$.
One of the popular approaches is the aforementioned Dynamic Mode Decomposition, which aims to find a matrix $A$ that evolves the system state as $\tilde{\bm{x}}(t+\Delta t) = A \bm{x}(t)$. Since its formulation in~\cite{schmid2010dynamic}, there have been numerous modifications to the method that, for example, involve different processing of the input data~\cite{kutz2016multiresolution} or apply Koopman operator theory to produce more accurate approximations~\cite{williams2015data}.
Another example is~\cite{pathak2017using,lu2018attractor}, where, in a reservoir computing framework, Lyapunov exponents for several chaotic processes
are estimated based only on the data. The obtained model is then used to reproduce the dynamics of the systems.
For Lorenz equations the prediction based on this approach starts to deviate from the original after $t \approx 7$, but there are no prediction errors reported~\cite{pathak2017using,lu2018attractor}.
We also mention~\cite{vlachas2018data} where, instead of reservoir computing, long short-term memory networks (LSTMs) are used to predict future states for high-dimensional chaotic systems.

To predict system evolution with neural network models, in this work
we consider another architecture, namely, residual networks (ResNets).
These networks are known to not suffer from the problem of vanishing and exploding gradients
and have been successfully applied to image classification problems.
We demonstrate their performance on three ODE systems, some of which are chaotic in nature, and compare it to the performance of a simple feedforward neural network.

An approach closely related to ResNet is recently employed for predicting dynamical system solution in \cite{pawar2019deep}. In this paper the authors consider the Lorenz system, but do not report the obtained errors. A difference with our experiment setting is that the starting points are taken on the attractor, which makes it an easier test problem.

The rest of the paper is organized as follows.
Section~\ref{sec:nn_concepts} gives a short overview of the neural network theory, provides details on residual networks and
describes the architectures we use to predict the system evolution.
Section~\ref{sec:dynsys} describes the dynamical systems of our interest.
Section~\ref{sec:experiment} outlines the settings for the experiments.
Experimental results are described in Section~\ref{sec:results},
while in Section~\ref{sec:future} we make conclusions and discuss possible directions for future work.

\section{Basic concepts of neural networks}
\label{sec:nn_concepts}

In this section we provide the background for neural networks, discuss their main principles and describe a special type of networks, ResNet, that
is used for our experiments.

\subsection{Neural networks and their architecture}
\label{sec:nn_arc_new}

Although neural networks have become a fundamental concept in deep learning, their definition is rather complicated and could be given in different ways. Oftentimes, they are described by their main properties and structure. In this article we consider \emph{feedforward neural networks} or \emph{multilayer perceptrons} (MLPs), a specific class of networks. According to \cite{goodfellow2016deep}, feedforward neural networks are models that construct an approximation $g(\bm{x}, \theta)$ to some function $\bar{g}(\bm{x})$ and learn the parameter values $\theta$ which result in the best function approximation.
Inspired by analogies from neuroscience and human brain, these models are nowadays present in many areas of science and industry.
Their structure can be described as follows.

The main unit of a feedforward neural network is a \emph{neuron}. It has a set of weights $\bm{w} = (w_0, w_1, \ldots, w_N)^T$. A neuron takes an $N$-dimensional vector $\bm{x} \in \mathbb{R}^N$ as an input, computes a linear combination of the
vector
components and applies some function $G$ to it. The output of the neuron is $y = G(w_0 + w_1 x_1 + \ldots + w_N x_N) = G(\bm{w}^T \hat{\bm{x}})$, where $\hat{\bm{x}} = (1, x_1, \ldots, x_N)^T \in \mathbb{R}^{N+1}$.
Throughout this paper we use the hat notation for vectors with an appended unit component.

Neurons are grouped into layers. A \emph{fully connected}, or \emph{dense}, \emph{layer} of $K$ neurons takes the same $\bm{x}$ as an input and computes a vector of $K$ neuron outputs: $\bm{y} = (G(\bm{w}^{1T} \hat{\bm{x}}), \ldots, G(\bm{w}^{KT} \hat{\bm{x}})^T)$, where $\bm{w}^{i} = (w^i_0, w^i_1, \ldots, w^i_N)^T,\ i=1,\ldots,K,$ are weights of the $i$-th neuron. $G$ is called \emph{an activation function} and is
taken to be the same for all neurons in a layer, while their weights are independent.
For simplicity, the equation can be rewritten as $\bm{y} = G(W \hat{\bm{x}})$, where $W = (\bm{w}^1 | \bm{w}^2 | \ldots | \bm{w}^K)^T \in \mathbb{R}^{K \times (N+1)}$ is a matrix of layer weights and $G$ is applied componentwise.

Neural network layers can also be based on other transformations of the input data, like convolution, pooling, etc.
Another notion relevant to our work, namely, a batch normalization layer, is discussed later in this article.

Finally, the layers are combined into a neural network. The important property of a feedforward neural network is that it
does not contain cycles, i.e., the outputs of the layers are not fed back into themselves \cite{goodfellow2016deep}.
Let us consider an example of a feedforward neural network with $L$ fully connected layers. Its first layer takes $\bm{x} \in \mathbb{R}^N$ and
gives a vector $\bm{G}_1 = G_1(W_1 \hat{\bm{x}})$. The second layer takes $\bm{G}_1$ as an input and returns $\bm{G}_2 = G_2(W_2 \hat{\bm{G}}_1)$.

This process continues
recursively as
$\bm{G}_{i+1} = G_{i+1}(W_{i+1} \hat{\bm{G}}_{i})$
and the network output is the output of its final layer $\bm{y}=G_L(W_L \hat{\bm{G}}_{L-1})$.

In this case, the parameters $\theta$ that need to be tuned in order to make the best approximation of $\bar{g}$ are the weights of the network $W_1$, $\ldots$, $W_L$, number of layers, size of each layer, activation functions etc. The architecture of the network, meaning a particular set of the parameters besides the weights, depends on the problem that needs to be solved. Intuitively, the bigger the problem, the more layers of larger size are used.
Activations $G_i$ are usually chosen to be nonlinear so that the network is able to approximate more complex functions.

Their properties are discussed in detail in \cite{ding2018activation, nwankpa2018activation}.
Here some examples of frequently used functions are given:

\begin{itemize}
    \item Sigmoid function $G(x) = 1/(1 + e^{-x})$;
    \item Hyperbolic tangent $G(x) = \tanh{x} = (e^x - e^{-x}) / (e^x + e^{-x})$;
    \item Rectified linear unit (ReLU) $G(x) = \max(0,x)$.
\end{itemize}

The ability of the neural networks to approximate a wide variety of functions $\bar{g}(\bm{x})$ is guaranteed by the \emph{universal approximation theorem}.
It has been reformulated and extended several times \cite{hornik1989multilayer, cybenko1989approximation, hornik1991approximation, lu2017expressive},
and the main result we need for this article is
that feedforward neural networks with finite number of neurons
can approximate continuous functions on a compact set in $\mathbb{R}^N$ with an arbitrary precision.

\subsection{Training and overfitting}
\label{sec:nn_train}

Once the architecture of the network is fixed, the weights are adjusted using optimization methods. A special \emph{loss function} $\mathcal{L}(W)$
is constructed that shows how ``close" a neural network approximation $g$ with a set of weights $W$ is to the true function $\bar{g}$ based on their values in points from a set $X=\{\bm{x}^1,\ldots,\bm{x}^M\}$. Typical example of such function is mean squared error (MSE) $\mathcal{L} = \frac{1}{M} \sum\limits_{i=1}^{M} \|g(\bm{x}^i, \theta) - \bar{g}(\bm{x}^i)\|^2$, where parameters $\theta$ include the weights $W$.
The process of adjusting network
weights via minimization of the loss function over $W$ is called \emph{training} of the network. Neural networks are trained using stochastic gradient descent, batch gradient descent and their modifications~\cite{bottou2018optimization, kiefer1952stochastic}. Such algorithms work either with the full dataset, or with its parts
(\emph{mini-batches}), or with a single sample from it to estimate the loss function gradient. The weights are updated via the algorithm called \emph{backpropagation}: due to the special structure of $g$, there is an efficient way to compute the gradients of loss function using chain rule \cite{goodfellow2016deep}.

The set of samples $X_{\text{train}}$ used in the loss function evaluation during the training is called \emph{train set}. Sometimes, we may observe high accuracy of trained neural networks on $X_{\text{train}}$, but significant drop in performance on other samples outside of $X_{\text{train}}$. It happens because the network has only learned to map the inputs from the train set to correct outputs, but hasn't gained the ability to generalize, i.e., perform well on previously unseen data. Such a phenomenon is called \emph{overfitting}~\cite{caruana2001overfitting, hawkins2004problem}.

A well known example of this phenomenon is fitting an $n$-th degree polynomial to given data points. If $n$ is very large, the resulting polynomial will be very close to the true values in the sampled points but its extrapolation beyond the fitted data will not yield a good accuracy. On the other hand, polynomials of a smaller degree may not be very accurate in the data points but will have better extrapolating qualities.

Possible solutions to this problem involve evaluating loss function not only on a train set, but also
on another set, called \emph{a test set} $X_{\text{test}}$,
of samples not present in $X$.
Monitoring the value of $\mathcal{L}(W)$ on the test set during the
training might help to
detect the moment of overfitting,
stop the training just before it takes place
and pick the best weights in terms of error from the previous training iterations (\emph{epochs}).
Such a technique is called \emph{early stopping}~\cite{caruana2001overfitting, raskutti2014early, yao2007early}.
The proportion between train and test set sizes, based on heuristics, is usually $80/20$ or $70/30$. Sometimes, to provide a final estimate of
the model
performance an additional \emph{holdout set} is used,
while the error during the training is measured on the test set.

In general, ways to avoid overfitting are called \emph{regularization}. It is defined in~\cite{goodfellow2016deep} as ``any modification to a learning algorithm that is intended to reduce its generalization error but not its training error". These modifications may
include additional constraints on parameter values and extra terms in the loss function. In this article, we apply the latter strategy.
By using a loss function
$$\mathcal{L_{\text{L2}}}(W) = \mathcal{L}(W) + \lambda \sum\limits_{i=1}^{L} \sum\limits_{j=0}^{K_i} \|W_i^j\|^2$$
with an additional term, where $\lambda$ is a small number, say $\texttt{1e-8}$, $L$ is a number of layers,
and $K_i$ is a size of $i$-th layer, we penalize the model
complexity
and enforce restrictions on the network weights by making them ``not too large".
Such a technique is called \emph{$L_2$ regularization} (and is also known as weight decay or Tikhonov regularization).  It often results in
a simpler model with better generalization properties.

Another technique
we use in our experiments is \emph{batch normalization}. In~\cite{ioffe2015batch}, the authors argue that it helps with the problem of covariate shift. Suppose that a
neural network is trained on images to recognize cats from dogs, but the set of cat pictures is imbalanced: it contains only black cats.
After training on such a 
dataset, the performance of the network on non-black cats will be poor. In other words, the images from train and test set come from different data distributions. 
This phenomenon is called \emph{the covariate shift problem}.

A
logical workaround for this situation would probably be
making the data more balanced by adding pictures of cats with different colors.
However, this may not necessarily help as this would change only the distribution of the input layer. The hidden (intermediate) layers may still suffer from internal covariate shift because the distribution of their inputs can change dramatically with each weight update.
This results in slow training and requires applying small learning rates.

A known solution to this problem, batch normalization, normalizes each layer's input over a mini-batch, thus reducing internal covariate shift.
For a mini-batch $B=\{\bm{x}_1, \ldots, \bm{x}_m\} \subset \mathbb{R}^N$ we compute the 
mean $\bm{\mu}_B = \frac{1}{m} \sum_{i=1}^m \bm{x}_i$ and the 
variance $\bm{\sigma}_B^2 = \frac{1}{m} \sum_{i=1}^{m} (\bm{x}_i - \bm{\mu}_B)^2$.
Then, the mini-batch is normalized as 
$\tilde{\bm{x}}_i = (\bm{x}_i - \bm{\mu}_B)/\sqrt{\bm{\sigma}_B^2 + \varepsilon}$, where $\varepsilon$ is a
small constant, division and root operations are applied componentwise. Finally, we apply scale and shift operations $\bm{y}_i = \bm{\gamma} \tilde{\bm{x}}_i + \bm{\beta}$, where $\bm{\gamma} \in \mathbb{R}^N$ and $\bm{\beta} \in \mathbb{R}^N$ are learnable parameters and multiplication is also performed componentwise, to obtain the layer output.
Such a technique reparametrizes layer outputs of deep networks so that the effect of the previous layers on the current layer output is decreased and the internal covariate shift is reduced. It also has a slight regularizing effect on the model because scaling on the mini-batch adds some noise to the output.

\subsection{Residual networks}
\label{sec:resnet}

In this subsection we discuss residual network (ResNet), an architecture that helps us to learn the dynamics of ODE systems. These types of networks are designed to combat vanishing and exploding gradient problem that occurs in very deep architectures.
More specifically, if a network has a large number of hidden layers, computing gradients via the backpropagation may result in very big or very small
values
and, hence, in inefficient weight updates~\cite{glorot2010understanding}. To overcome this problem, \cite{he2016deep}~offers to
replace the multiplications with summations.
The main idea is to use ``blocks'' of layers that will learn the difference between the input and the desired output and then just add the output of the block to the input data.

Consider a ResNet block of the following structure:
\begin{equation}
\label{RNblock}
FC \to BN \to \textit{ReLU} \to FC,
\end{equation}
where $FC$ is a fully connected layer, $BN$ denotes batch normalization, and
$\textit{ReLU}$ is a rectified linear unit function applied componentwise.
If $\bm{x}_n$ is the block input and $G(\bm{x}_n)$ is its output (i.e., the output of the second fully connected layer), then the total output is defined as
$$
\bm{x}_{n+1} = \bm{x}_n + G(\bm{x}_n).
$$
This value is then passed to the next block
of the same structure. As a result, forward propagation through the neural network with $M$ such blocks can be written as
$$
\bm{x}_{M} = \bm{x}_1 + \sum\limits_{i=1}^{M-1}G(\bm{x}_i).
$$
This last relation defines our neural network approximation
$$
\tilde{\bm{x}}(t+\Delta t) = g(\bm{x}(t), \theta),
$$
with $\bm{x}_1=\bm{x}(t)$, $\bm{x}_M=\tilde{\bm{x}}(t+\Delta t)$
and $\theta$ including the discussed neural network
architecture and the layer weights.
This approach, as mentioned earlier, reduces the amount of multiplications required for training and evaluating gradients via backpropagation and, thus, the gradients are less likely to vanish or explode.

Although there is no mention of numerical methods in the original article,
other papers~\cite{weinan2017proposal,lu2017beyond,chen2018neural}
have rightfully noted that the structure of the network resembles the
classical explicit Euler method for solving differential equations.
Indeed, for the initial-value problem
$$
\dot{\bm{x}}(t) = F(\bm{x}(t)), \qquad
\bm{x}(t_0) = \bm{x}_0,
$$
solution in the explicit Euler scheme at time moments $t_n = t_0 + nh$
can be found via updates
$$
\bm{x}_{n+1} = \bm{x}_n + h F(\bm{x}_n),
$$
where $h>0$ is a step size.
This
is one of our motivations for using ResNet to learn dynamics of the ODE systems.
We apply several neural networks based on this idea. They differ in the number of layers, their size and depth.
It should be noted that the aim of the article is not to find an optimal
neural network structure that would result in the smallest prediction error, but to evaluate different networks and see whether they are capable to predict the evolution.
We use the following four architectures:
\begin{itemize}
    \item \textbf{RN1}: the simplest of the architectures. It consists of 4~ResNet blocks. Each ResNet block has structure~\eqref{RNblock} and its output is added to the block's input.
    The size of each fully-connected layer is $10$.
    \item \textbf{RN2}: has the same structure as RN1
    (cf.~\eqref{RNblock}), except that the size of fully connected layer is increased to $50$.
    \item \textbf{RN3}: a deeper network with 6 ResNet blocks and FC size of $50$.
    \item \textbf{RN4}: a network with a deeper structure of ResNet block: $FC \to BN \to ReLU \to FC \to BN \to ReLU \to FC$. Like the first two networks, it has 4 blocks
    but the size of fully connected layer is 15.
\end{itemize}
We also compare the performance of the ResNet architectures to
an MLP architecture $FC \to BN \to ReLU \to FC \to BN \to ReLU \to FC \to BN \to ReLU \to FC \to Sigmoid$ on one of the ODE systems.
For all the described architectures the input is normalized
to~$[0,1]$.

\section{Dynamical systems}
\label{sec:dynsys}

In this section, the dynamical systems are described on which
our neural network models are tested.
These systems are: the Van der Pol oscillator,
the Lorenz system and the R\"ossler system.
The description of these ODE systems, as well as their governing equations, is given below. In addition, Table~\ref{tab:params} shows the parameters of the equations considered for the experiments.

\subsection{Van der Pol oscillator}
\label{sec:e0201}

For the oscillator described by B.~van~der~Pol in~\cite{van1926lxxxviii}, the original equation can be written as
$$\ddot{x} - \mu (1 - x^2) \dot{x} + x = 0,$$
where the usual meaning of $x(t)$ is position (but this may
differ depending on the application) and $\mu$ is a scalar parameter that characterizes damping.
In the experiments, we set its value to $\mu = 3$.
If the time derivative $y(t)=\dot{x}(t)$ is introduced as additional variable, the equation can be reduced to a system of first order differential equations $\dot{\bm{x}}=F(\bm{x})$,
with $\bm{x}(t)=[ x(t), y(t)]^T$:
\begin{equation*}
\begin{aligned}
\dot{x}(t) &= y,\\
\dot{y}(t) &= \mu (1 - x^2) y - x.
\end{aligned}
\end{equation*}
The limit-cycle orbits for this system in the phase space are shown in Figure~\ref{ds_vdp}.

The system serves as an example of rather ``simple" dynamics, so it makes sense to test our models on such problem first.
Before feeding the system data to the neural networks, we scale it by the maximum absolute value among all the vector components from training and testing data, so that each solution component lies in $[0,1]$.

\subsection{Lorenz system}
\label{sec:e0202}

\begin{figure}
\centering
\begin{subfigure}[b]{.49\linewidth}
\includegraphics[width=\linewidth]{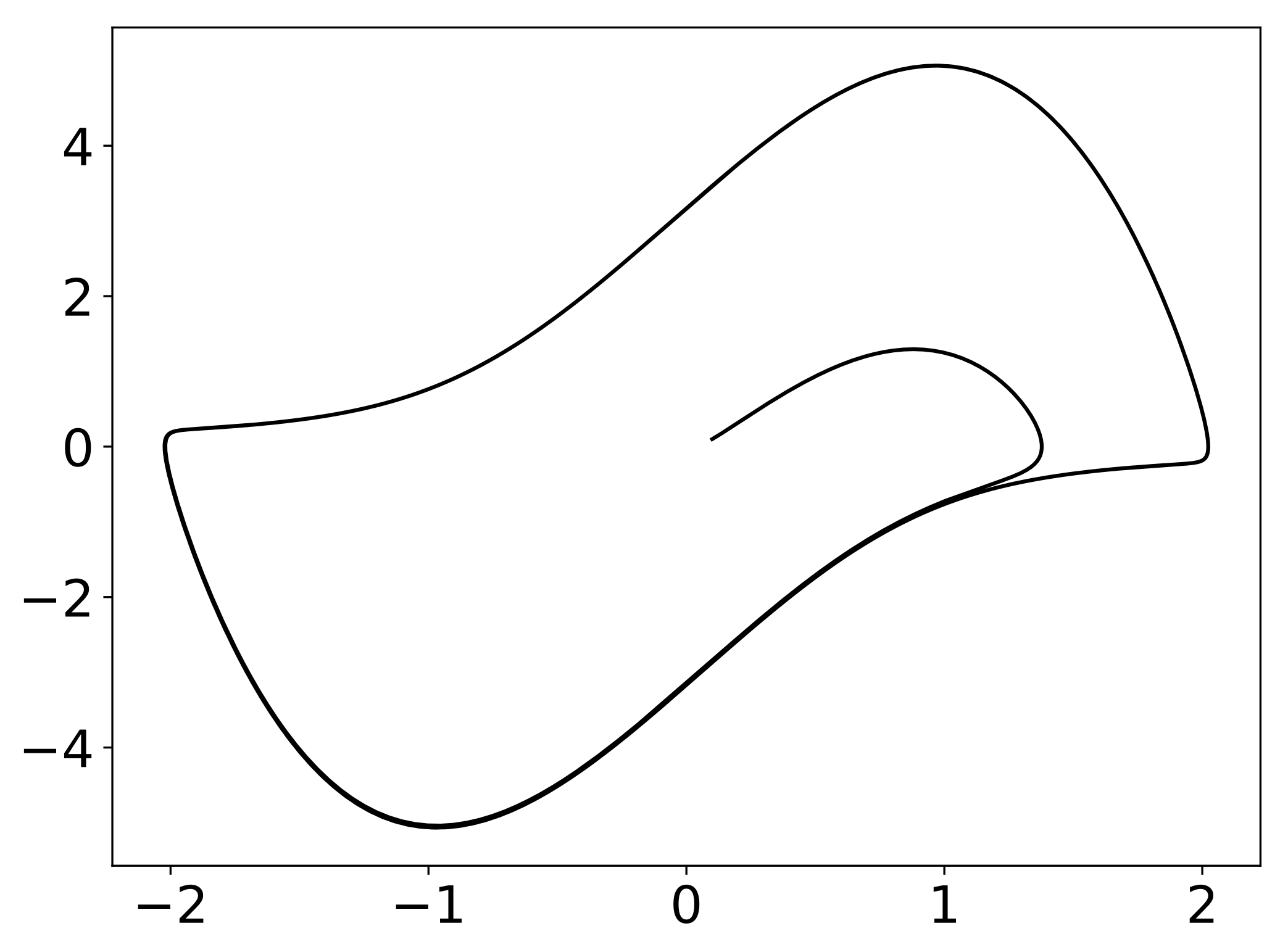}
\caption{ }\label{ds_vdp}
\end{subfigure}

\begin{subfigure}[b]{.49\linewidth}
\includegraphics[width=\linewidth]{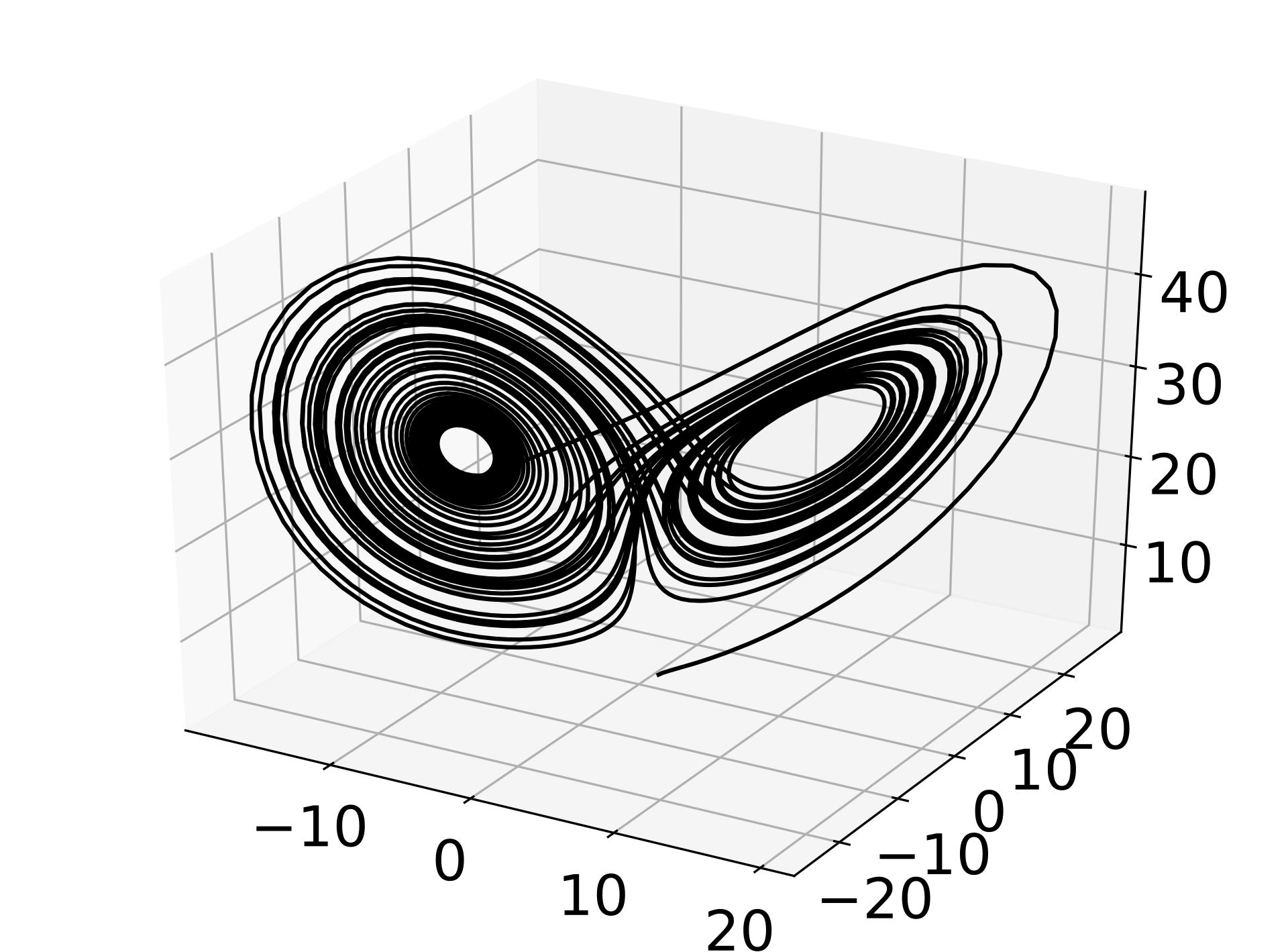}
\caption{ }\label{ds_lorenz}
\end{subfigure}
\begin{subfigure}[b]{.49\linewidth}
\includegraphics[width=\linewidth]{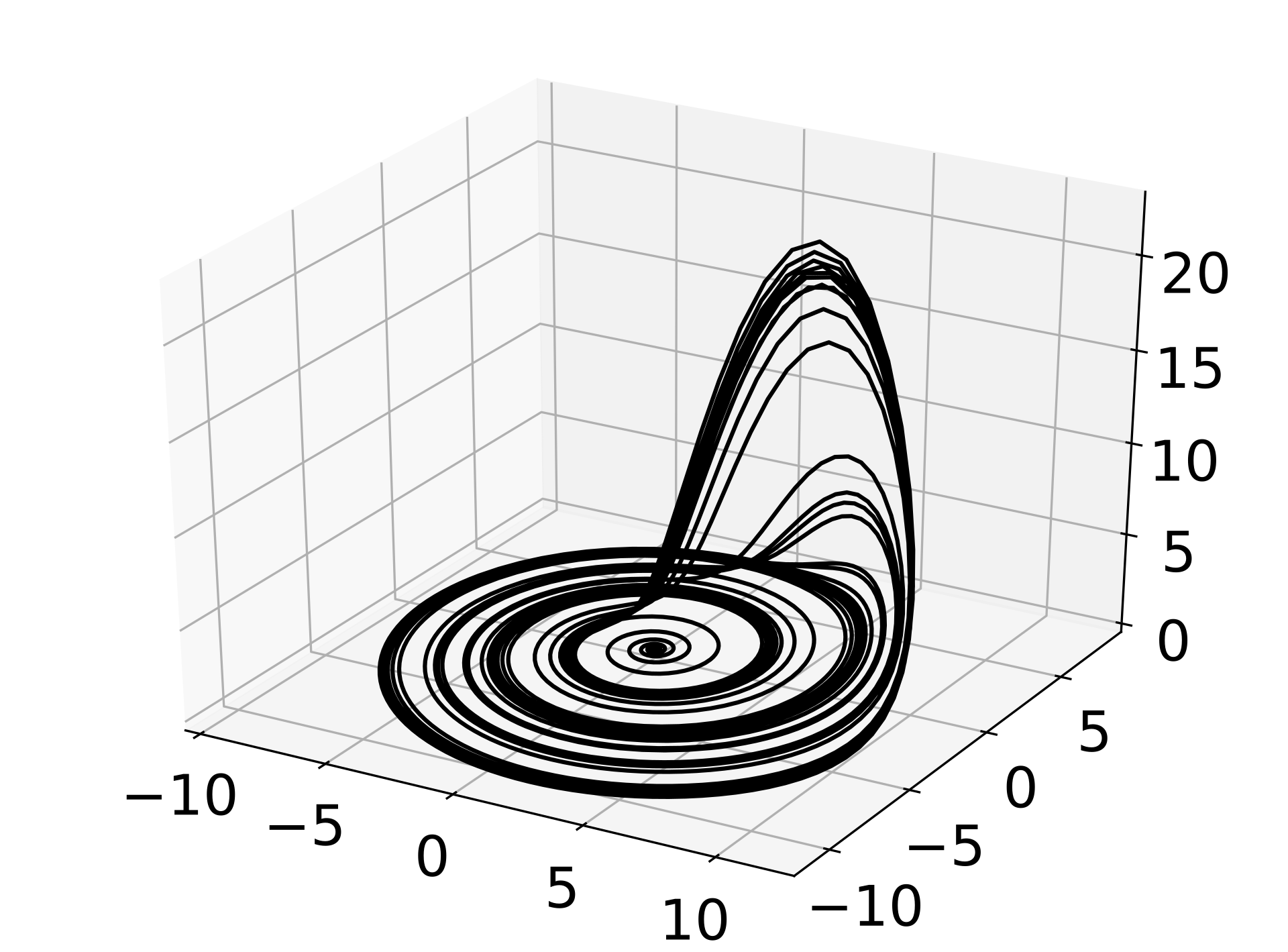}
\caption{ }\label{ds_rossler}
\end{subfigure}
\caption{Trajectories of the dynamical system solutions: (a)~Van der Pol oscillator, (b)~Lorenz system, (c)~R\"ossler system}
\label{ds}
\end{figure}

Denoting $\bm{x}(t)=[ x(t), y(t), z(t)]^T$, we write
the Lorenz ODE system $\dot{\bm{x}}=F(\bm{x})$ componentwise as
\begin{equation*}
\begin{aligned}
\dot{x} &= \sigma (y-x),
\\
\dot{y} &= x (\rho - z) - y,
\\
\dot{z} &= xy - \beta z,
\end{aligned}
\end{equation*}
where we set $\sigma = 10$, $r=28$, $b=8/3$.
This ODE system is described by E.~Lorenz in \cite{lorenz1963deterministic}. It exhibits a chaotic behavior and
has a butterfly-shaped solution set, called attractor (Figure~\ref{ds_lorenz}).
For this system we scale the system data for feeding the neural networks
in a different way than it is done for the other two systems.
Instead of scaling by the maximum values of the training and testing
vector entries, we scale the data by the theoretically known
solution bounds~\cite{yu2010estimating}.
Denote
$$m = \sigma + \rho,$$

$$r_1 = \sqrt{\frac{-\beta m^2}{4 \max \{-\sigma, -1, -\beta\}}},$$

$$r_2^2 =\begin{cases}
\dfrac{\beta^2 m^2}{4 \sigma (\beta - \sigma)},& \text{if } \sigma \leq 1, \sigma \leq \beta/2 \\
\dfrac{\beta^2 m^2}{4 (\beta - 1)},& \text{if } 1 < \sigma, 1 \leq \beta / 2 \\ m^2, & \text{if } \sigma > \beta/2, 1 > \beta/2 \end{cases},$$

$$R = \max \{r_1 + |m| / 2, r_2\}.$$
The following result holds~\cite{yu2010estimating}.
If the initial conditions lie in the sphere
$$
\Omega = \{(x,y,z)\ |\ x^2 + y^2 + (z-m)^2 \leq R^2\},
$$
then the evolving system trajectory is contained in $\Omega$, too~\cite{yu2010estimating}.
In our case $m = 38$ and $R \approx 50.0$.

\subsection{R\"ossler system}
\label{sec:rossler}

Here, our test ODE system $\dot{\bm{x}}=F(\bm{x})$ is
taken from paper~\cite{rossler1976equation}
published in 1976 by O.~R\"ossler.
There he introduced an ODE system
\begin{equation*}
\begin{aligned}
\dot{x} &= - y - z,
\\
\dot{y} &= x + a y,
\\
\dot{z} &= b + z (x - c),
\end{aligned}
\end{equation*}
with $a$, $b$, $c$ being given real parameters.
We use the same parameter values as in the original paper
of R\"ossler: $a = 0.2$, $b = 0.2$, $c = 5.7$.
Similar to the Lorenz system, the system exhibits a chaotic dynamics. Figure~\ref{ds_rossler} shows the attractor of the R\"ossler system.

The R\"ossler and Lorenz system are examples of low-dimensional models with
rather
complex and chaotic dynamics.
Due to their chaotic nature, both of them are very sensitive to initial
conditions.
Hence, these problems provide a good challenge for our
neural network models.

\section{Experiment setting}
\label{sec:experiment}

In the experiments presented here we evaluate the four types of networks discussed on the solution samples of the three ODE systems and compare them to the MLP architecture.
The process of generating the data for each system is as follows.
First, $N = 12\,000$ points are sampled from $N(0,2I)$.
Note that, depending on the problem
dimension, these points are vectors sampled either from $\mathbb{R}^2$ or $\mathbb{R}^3$.
Then, the differential equations corresponding to the problem are solved
numerically with these points as initial conditions and $12\,000$ trajectories $\bm{x}_1(t)$, \dots, $\bm{x}_N(t)$ are obtained.
For this purpose the LSODA solver of the function \texttt{odeint}
from the Python's
SciPy library run with default tolerances
$\mathtt{atol}= \texttt{1.49012e-8}$, $\mathtt{rtol}= \texttt{1.49012e-8}$, is employed.
The solver uses an automatically chosen time step size and produces
solution at $2500$ time moments $0$, $\Delta t$, $2\Delta t$, \dots,
$T-\Delta t$, $T$,
with $T$ being  the final time and $\Delta t=T/2500$.
For each system, we choose $T$ such that
the points evolve around the attractor or limit cycle several times
to sufficiently capture the evolution of the system in the sampled data.
The parameter values
are given in Table~\ref{tab:params}. In the table, we also report the Lyapunov time $T_L$ for every dynamical system \cite{sprott2003chaos, sprott2010elegant}. Note that, since we consider a simple non-chaotic case of the
Van~der~Pol oscillator, its Lyapunov time is, formally speaking,
infinite, while the other two systems have finite Lyapunov times.

\begin{table}
\begin{center}
  \caption{Descriptions and numerical values for some of the parameters}
  \label{tab:params}
  \begin{tabular}{| l | l | l | l | l |}
    \hline
         Parameter & Description & Van der Pol & Lorenz & R\"ossler \\ \hline
     --- & Equation parameters & $\mu=3$ & \makecell{$\sigma = 10$,\\ $r=28$,\\ $b=8/3$} & \makecell{$a = 0.2$,\\ $b = 0.2$,\\ $c = 5.7$}   \\ \hline
    $N$ & Number of points & $12\,000$ & $12\,000$ & $12\,000$   \\ \hline
    $N_1$ & \makecell[l]{Number of training\\ set samples} & $8\,000$ & $8\,000$ & $8\,000$   \\ \hline
    $N_2$ & \makecell[l]{Number of test\\ set samples} & $2\,000$ & $2\,000$ & $2\,000$   \\ \hline
    $N_3$ & \makecell[l]{Number of holdout\\ set samples} & $2\,000$ & $2\,000$ & $2\,000$   \\ \hline
    $T$ & Final time    & 25 & 25 & 125   \\ \hline
    $T_L$ & Lyapunov time    & $\infty$ & 1.1 & 14.0   \\ \hline
    $\Delta t=T/2500$ & Integration timestep  & 0.01  & 0.01 & 0.05  \\
    \hline
  \end{tabular}
\end{center}
\end{table}

Once the numerical data for experiments are obtained,
we evaluate the neural networks RN1--RN4 on it.
For each of the $12\,000$ test trajectories,
the solution samples are divided into the input--output pairs
$(\bm{x}(t), \bm{x}(t+\Delta t))$, for times $t=0, \ldots, T - \Delta t$.
This means that for each input $\bm{x}(t)$,
the neural network should produce an output $\tilde{\bm{x}}(t + \Delta t)$
close to the ODE solution $\bm{x}(t + \Delta t)$.
We note that each state is a vector comprised of coordinates of
$12\,000$ points, so its dimension is $2N=24\,000$ for the Van~der~Pol system
and $3N=36\,000$ for Lorenz and R\"ossler systems.

Out of the total $N$ trajectories, we consider
$N_1 = 8000$ trajectories for training the networks and
$N_2 = 2000$ trajectories for their testing.
The remaining $N_3 = N - N_1 - N_2 = 2000$
trajectories are a holdout set which we
use to evaluate the accuracy of predictions by the trained models.
The training algorithm is Adam \cite{kingma2014adam} with the mini-batches of size $2048$.

The mean squared error with L2 regularization, defined as
$$
\mathcal{L_{\text{L2}}} = \frac{1}{mn} \sum\limits_{i=1}^{n} \sum\limits_{j=1}^{m} \|\bm{x}_i(t_j) - \tilde{\bm{x}}_i(t_j)\|^2 + \lambda \sum\limits_{i=1}^{L} \sum\limits_{j=0}^{K_i} \|W_i^j\|^2,
$$
where $n$ is a number of trajectories in the set, $m$ is a number of considered pairs for each trajectory, regularization term notations follow Section~\ref{sec:nn_train}, $\lambda = 1e-10$, is used as a loss function. Note that the particular values of $n$ and $x_i$ depend on the set of our choice (train, test or holdout); the value of $m$ depends on the training step that we discuss later in the article.

Several techniques are used to speed up the training of the networks.
First,
only a part of the sampled $N_1$ trajectories is used: of all the input--output
pairs $(\bm{x}(t), \bm{x}(t+\Delta t))$, only every fifth pair is
used for training.
Second, we start with a small number of samples in train and test sets and then iteratively add more samples after several training epochs.
More specifically, let us denote the set of input training samples
by $\Xin$ and the set of expected output samples
by $\Xout$.
At the beginning of the training both sets are empty.
We define an $i$-th training step as the following sequence of actions.
\begin{enumerate}
    \item Add each fifth pair to the sets, i.e.,
    add $\bm{x}(5 (i-1))$ to $\Xin$ and $\bm{x}(5 (i-1) + 1)$ to $\Xout$.
    \item
    Take the weights from the previous training step and initialize the network with them (at the first step the network is initialized with random weights).
    \item Train the network on pairs from $\Xin$ and $\Xout$.
    Stop the training as soon as one of two events occurs: the network performs 5000 training epochs or its relative prediction error on the test set
    $$\varepsilon_{\text{i}} = \frac{1}{i N_2} \sum\limits_{j=1}^{N_2} \sum\limits_{k=1}^{i} \frac{\|\bm{x}_j(5(k-1) + 1) - \tilde{\bm{x}}_j(5(k-1) + 1)\|}{\|\bm{x}_j(5(k-1) + 1)\|},$$
    where summation is performed over evolution pairs from $N_2$ trajectories added during steps $1$ to $i$,
    $x_j$ are taken from the test set,
    is within a certain tolerance
    (in our experiments, $0.0015$).
    \\
    \item Save the resulting weights (they are used as initial values for the next step).
\end{enumerate}

This algorithm allows for a more efficient data usage and a faster training.
It should be stressed that the same sizes of train and test sets on each step (and, thus, the same complexity) could be obtained if we used a five times
greater step size $5 \Delta t$
and considered the whole dataset of the trajectories.
However, this would not be as beneficial as our procedure.
In this case the data would be less representative of
the system
evolution and the period of prediction for the networks would be five times longer.
The stopping criterion is motivated by the preliminary experiments with Lorenz system:
we observe that the error of prediction decreases after 5000 epochs down to
$\approx 0.0015$, which is an acceptable accuracy for us.

To
evaluate the predictions of
the four networks, we test
their performance in two cases:
\begin{itemize}
    \item Training the networks on samples for $t \in [0,T/2]$ and predicting the trajectories from the holdout set for $t \in [T/2,T]$.
    \item Training the networks on samples for $t \in [0,T/4]$ and predicting the trajectories from the holdout set for $t \in [T/4,T]$.
\end{itemize}
Training is performed according to the procedure described above. After training, both networks are given only one snapshot, $\bm{x}(T/4)$ or $\bm{x}(T/2)$, of samples from the holdout set. They iteratively predict the solution at the next time moment over a time step $\Delta t$ and use the result as an input to predict further in time. In both cases
we measure the average relative prediction error on the holdout set as $$\varepsilon_{\text{avg}} = \frac{1}{M N_3} \sum\limits_{i=1}^{N_3} \sum\limits_{j=1}^{M} \frac{\|\bm{x}_i(t_j) - \tilde{\bm{x}}_i(t_j)\|}{\|\bm{x}_i(t_j)\|},$$
where $M$ is the prediction step number ($M=250$ for $t \in [T/2,T]$ and $M = 375$ for $t \in [T/4,T]$) and
$t_j$ are the time moments from the prediction interval taken with
a time step $\Delta t$,
$x_i$ are taken from the holdout set.
We also measure the relative prediction error at $t=T$ defined as
$$
\varepsilon_T = \frac{1}{N_3} \sum\limits_{i=1}^{N_3} \frac{\|\bm{x}_i(T) - \tilde{\bm{x}}_i(T)\|}{\|\bm{x}_i(T)\|}.
$$
The same procedure is also carried out for an MLP network described in Section~\ref{sec:resnet}.

\section{Discussion of the results}
\label{sec:results}

    \begin{figure}
    \begin{subfigure}{.5\textwidth}
      \centering
      \includegraphics[width=.99\linewidth]{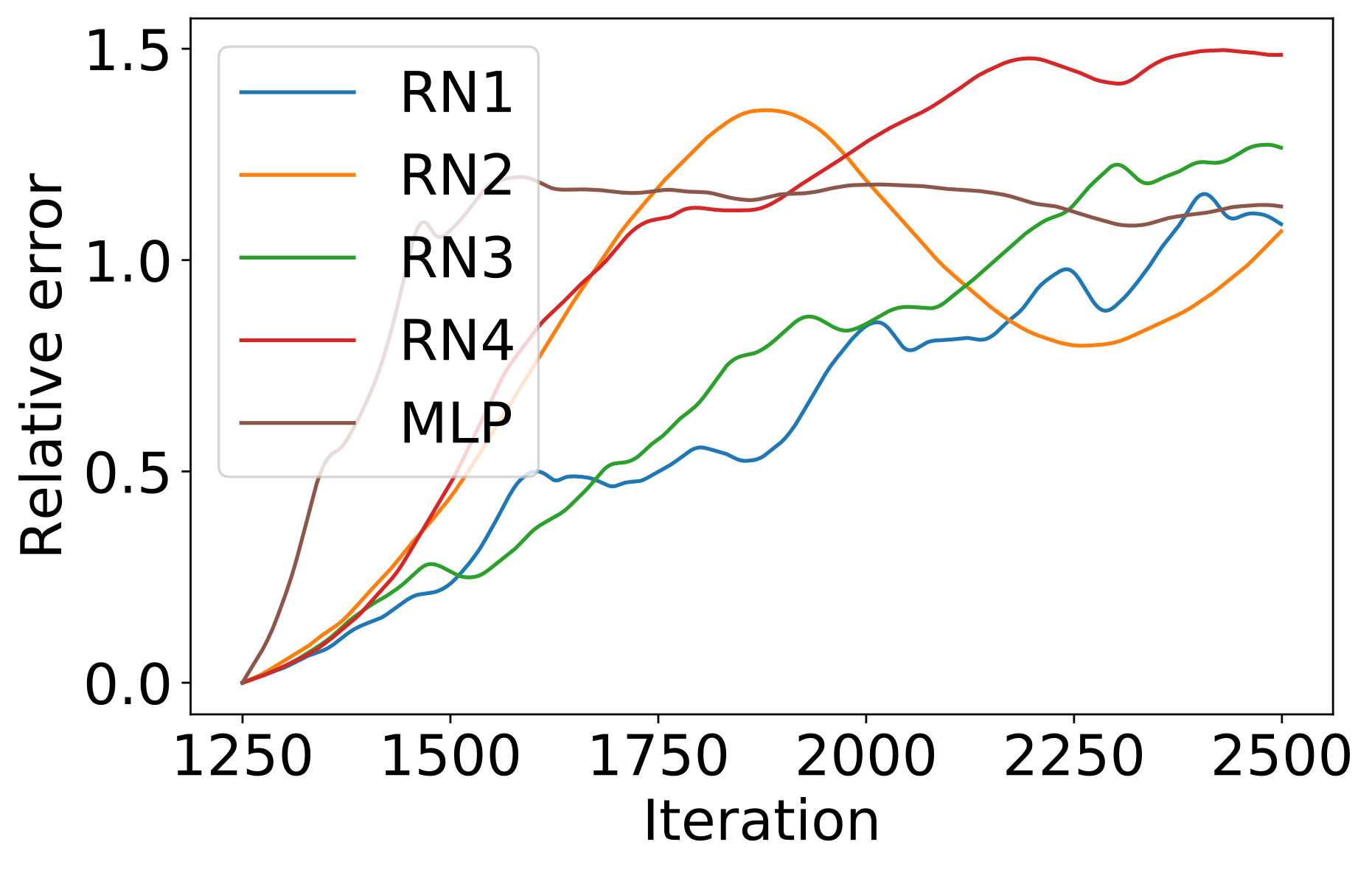}
      \caption{ }
      \label{vdp_250_1250_2500}
    \end{subfigure}%
    \begin{subfigure}{.5\textwidth}
      \centering
      \includegraphics[width=.99\linewidth]{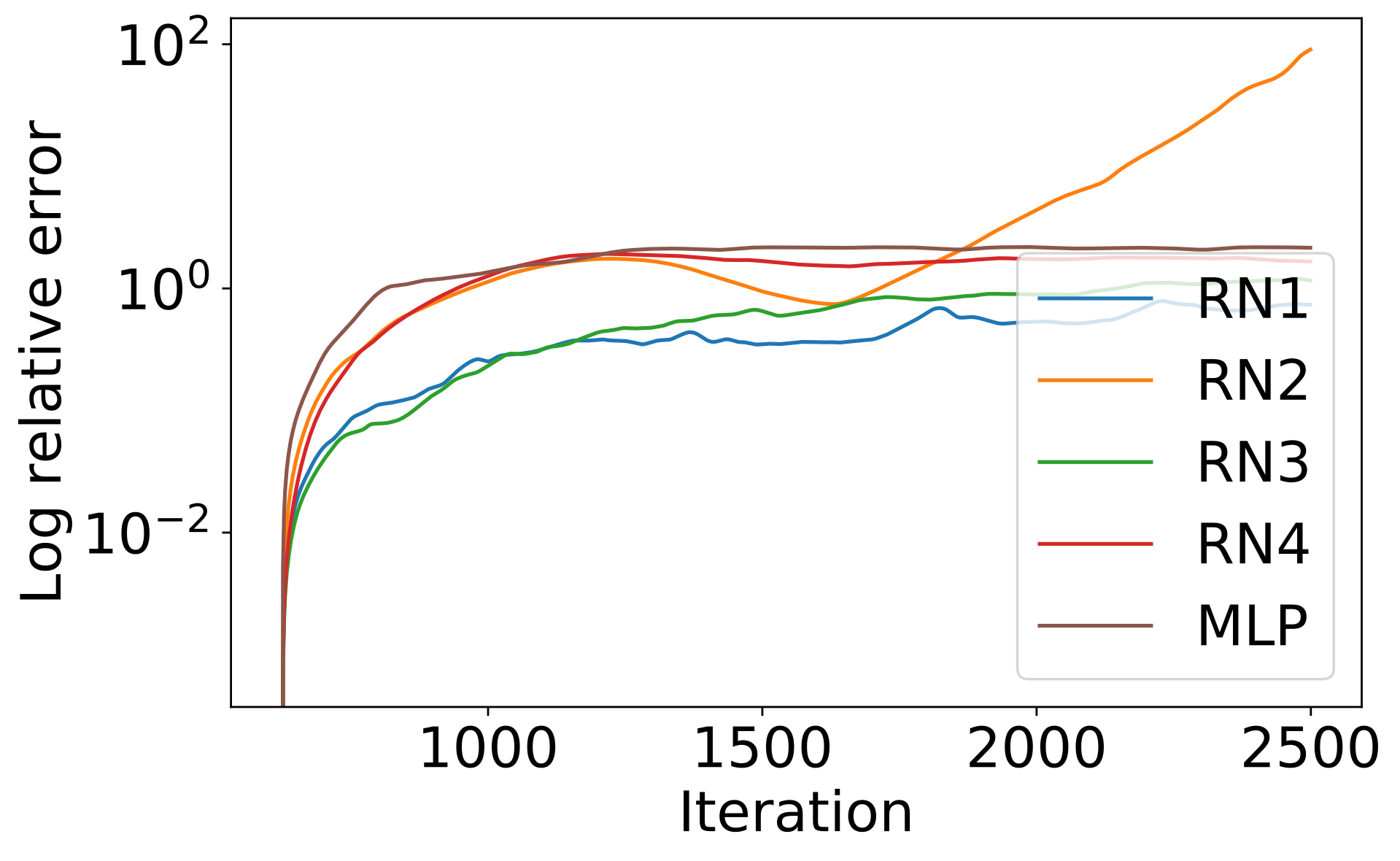}
      \caption{ }
      \label{vdp_125_625_2500}
    \end{subfigure}
    \caption{Stepwise ResNet and MLP prediction errors for Van der Pol oscillator problem from $T/2$ to $T$ (a) and from $T/4$ to $T$ (b)}
    \label{fig:vdp_errors}
    \end{figure}

\begin{figure}
\centering
\begin{subfigure}[b]{.49\linewidth}
\includegraphics[width=\linewidth]{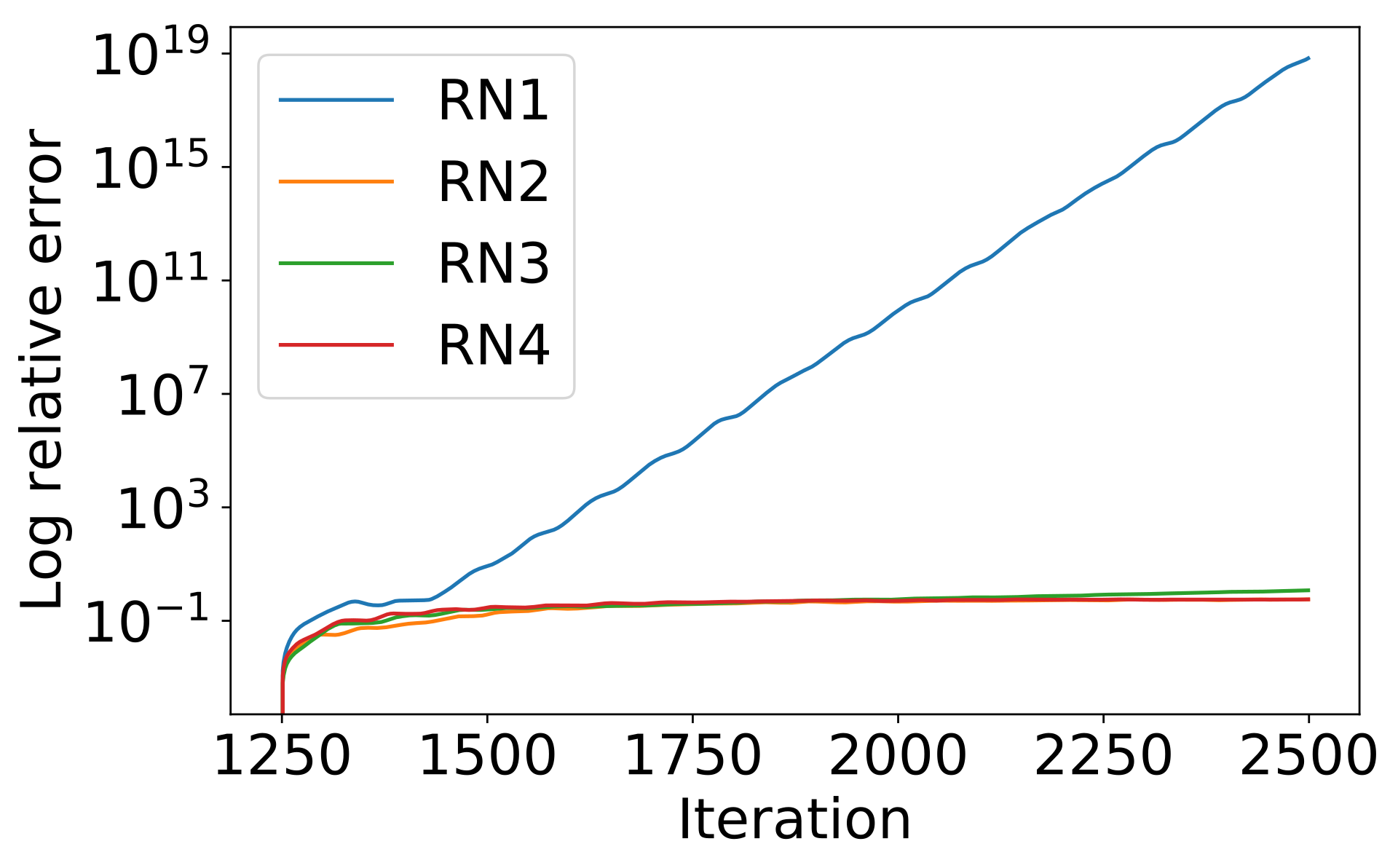}
\caption{ }\label{lorenz2_250_1250_2500_log}
\end{subfigure}
\begin{subfigure}[b]{.49\linewidth}
\includegraphics[width=\linewidth]{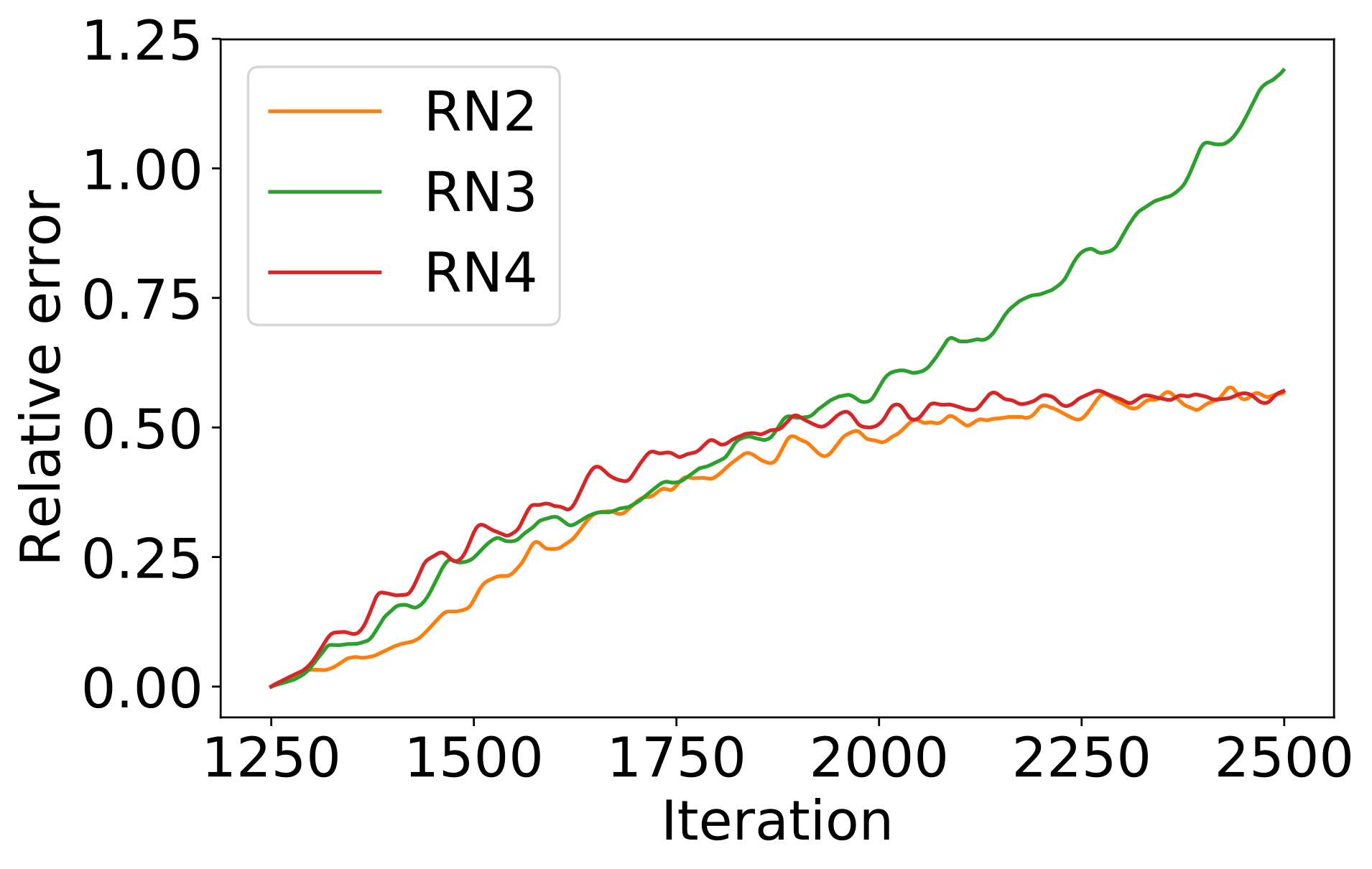}
\caption{ }\label{lorenz2_250_1250_2500}
\end{subfigure}
\begin{subfigure}[b]{.49\linewidth}
\includegraphics[width=\linewidth]{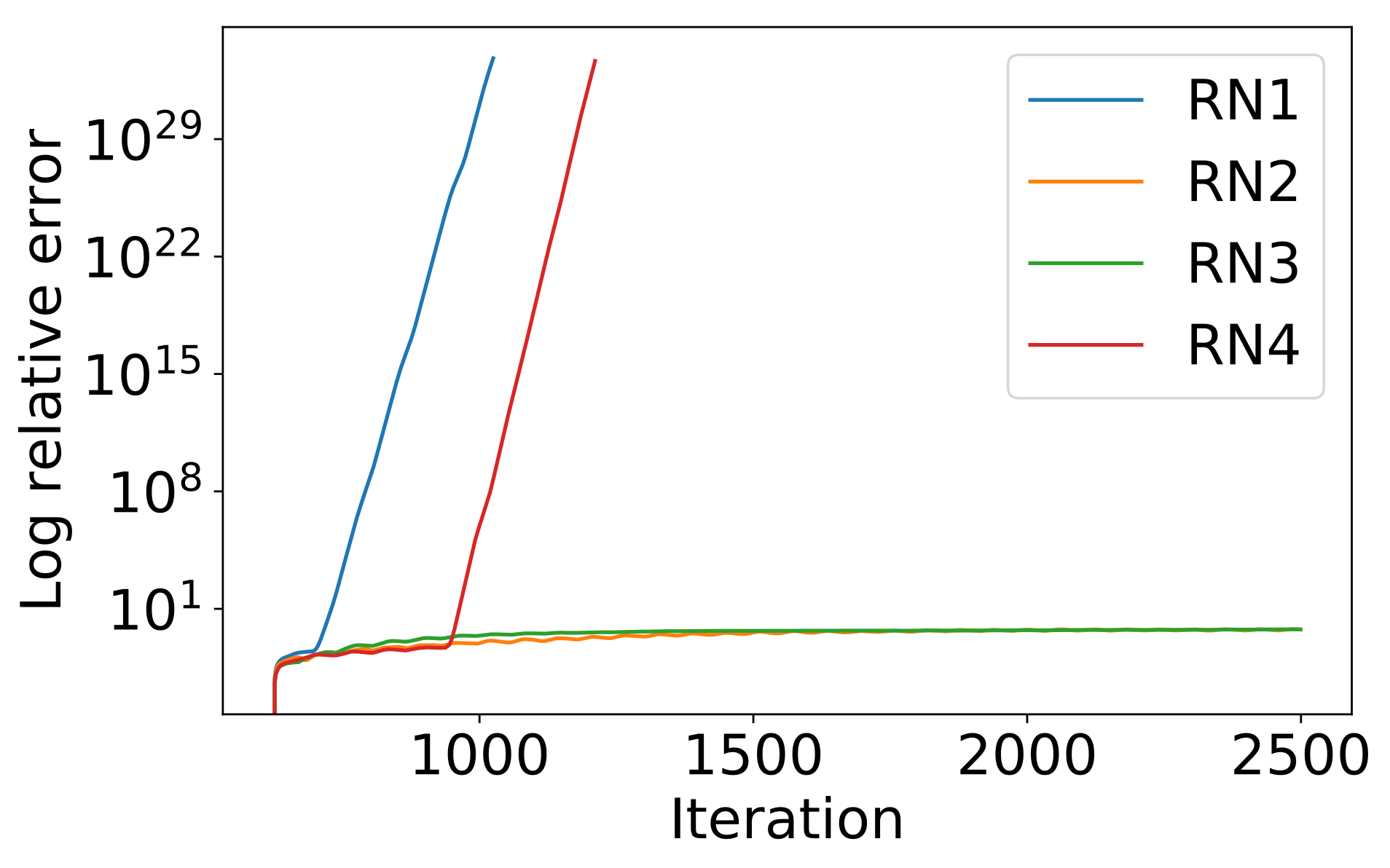}
\caption{ }\label{lorenz2_125_625_2500_log}
\end{subfigure}
\caption{Stepwise ResNet prediction errors for Lorenz system from $T/2$ to $T$ (a,b) and from $T/4$ to $T$ (c)}
\label{fig:lorenz_errors}
\end{figure}

    \begin{figure}
    \begin{subfigure}{.5\textwidth}
      \centering
      \includegraphics[width=.9\linewidth]{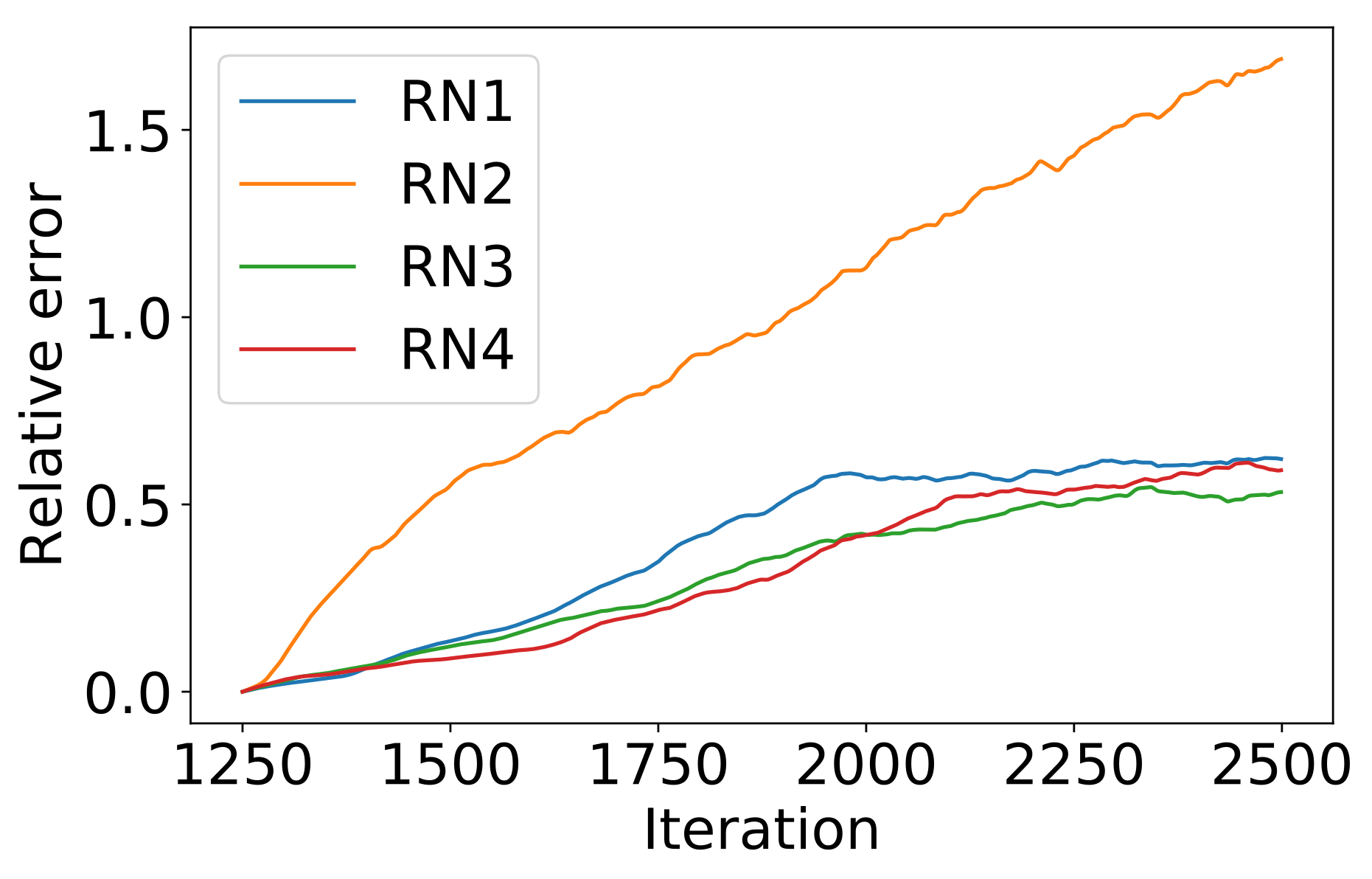}
      \caption{ }
      \label{rossler_250_1250_2500}
    \end{subfigure}%
    \begin{subfigure}{.5\textwidth}
      \centering
      \includegraphics[width=.9\linewidth]{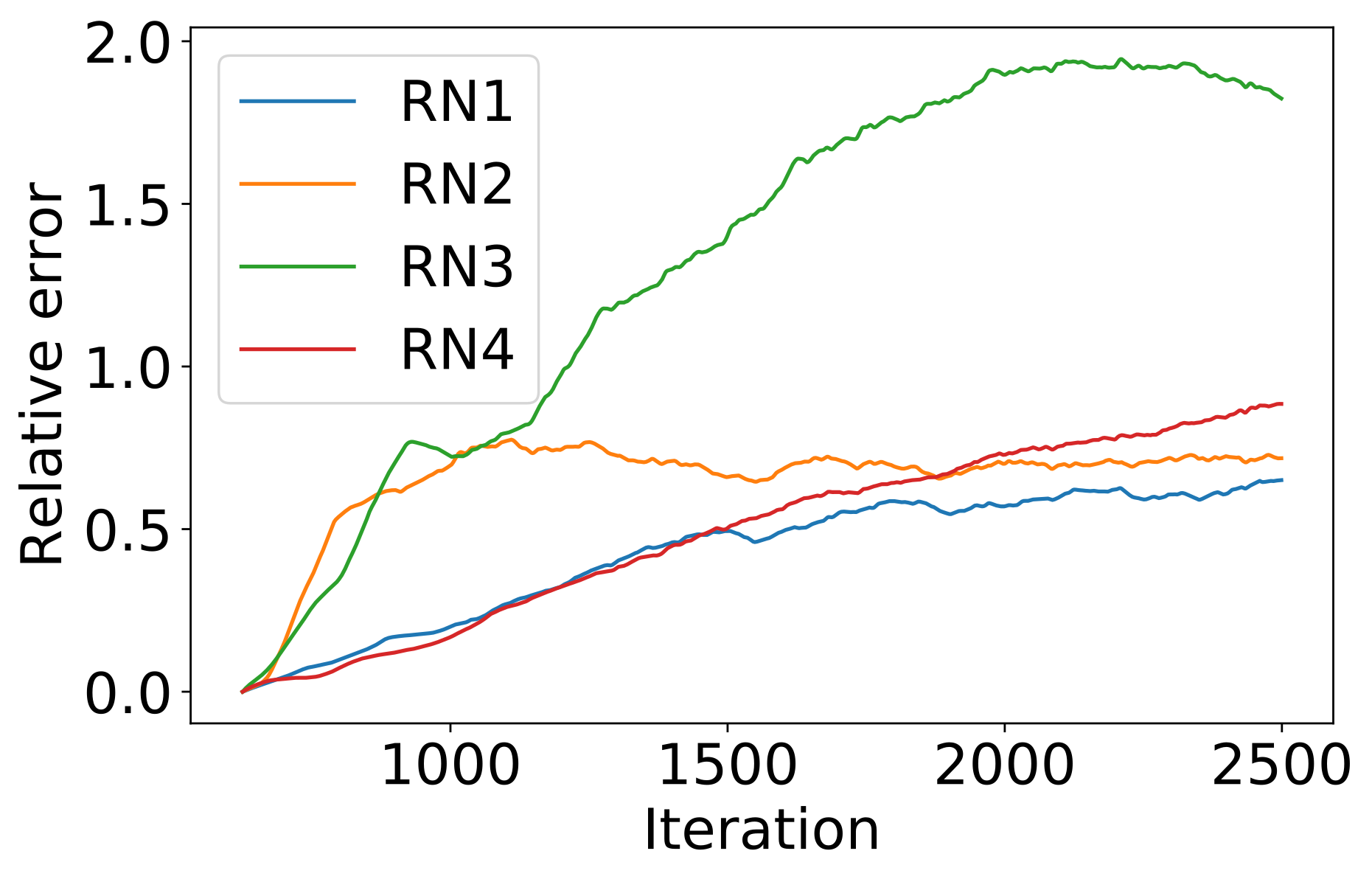}
      \caption{ }
      \label{rossler_125_625_2500}
    \end{subfigure}
    \caption{Stepwise ResNet prediction errors for R\"ossler system from $T/2$ to $T$ (a) and from $T/4$ to $T$ (b)}
    \label{fig:rossler_errors}
    \end{figure}

\begin{figure}
\centering
\begin{subfigure}[b]{.49\linewidth}
\includegraphics[width=\linewidth]{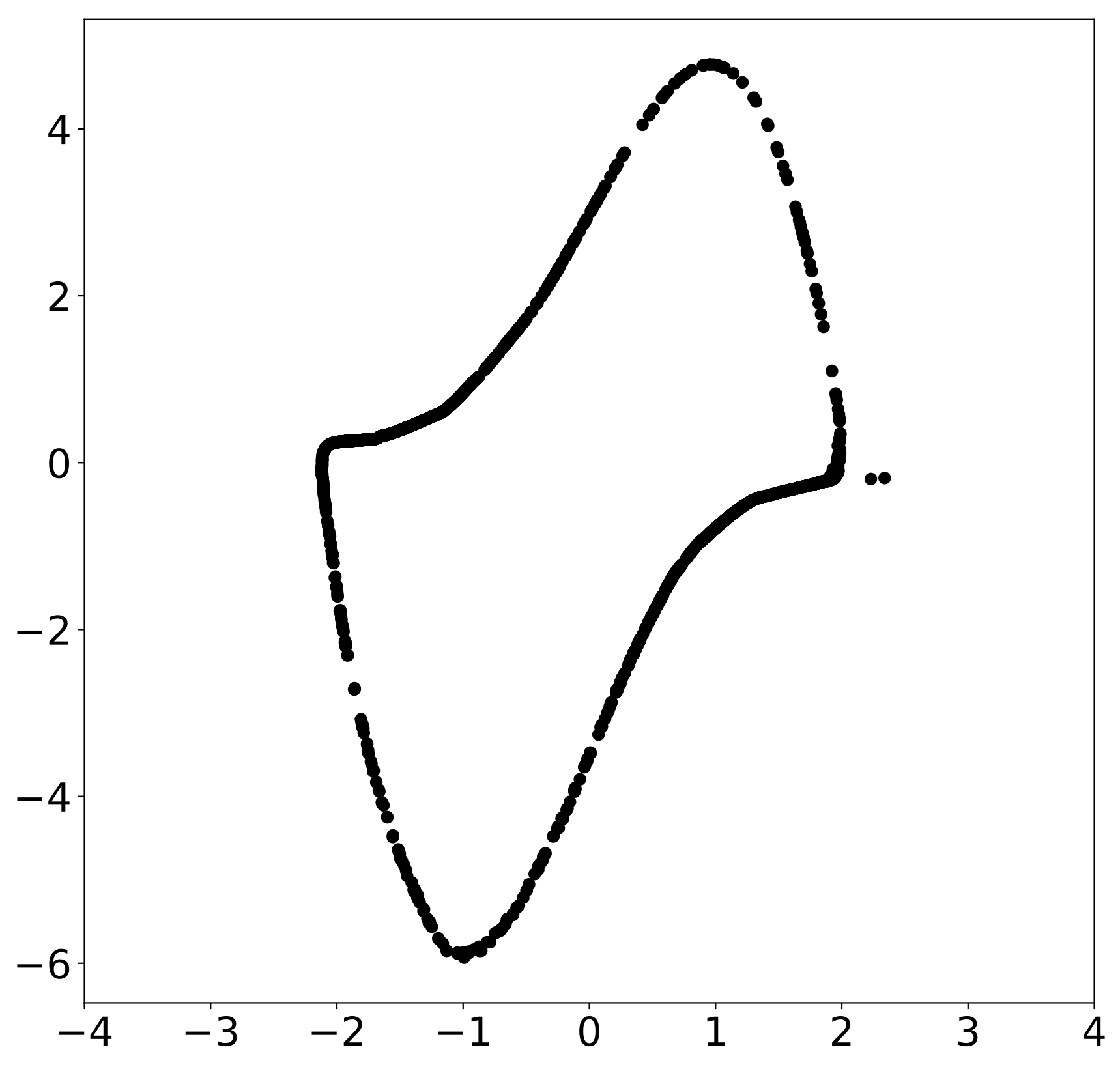}
\caption{ }\label{vdp_250_evol_1250_2500_pred}
\end{subfigure}
\begin{subfigure}[b]{.49\linewidth}
\includegraphics[width=\linewidth]{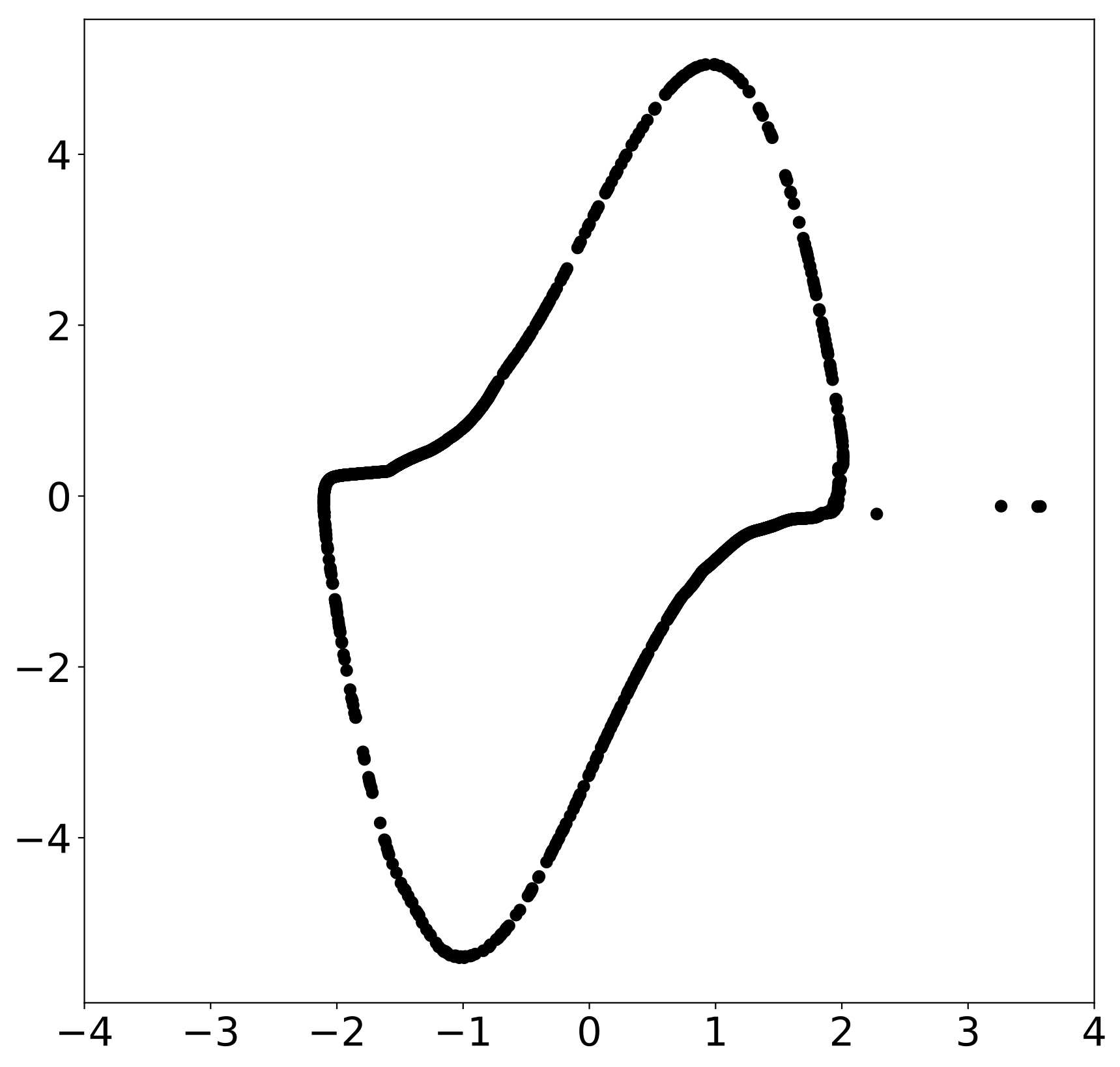}
\caption{ }\label{vdp_125_evol_625_2500_pred}
\end{subfigure}
\begin{subfigure}[b]{.49\linewidth}
\includegraphics[width=\linewidth]{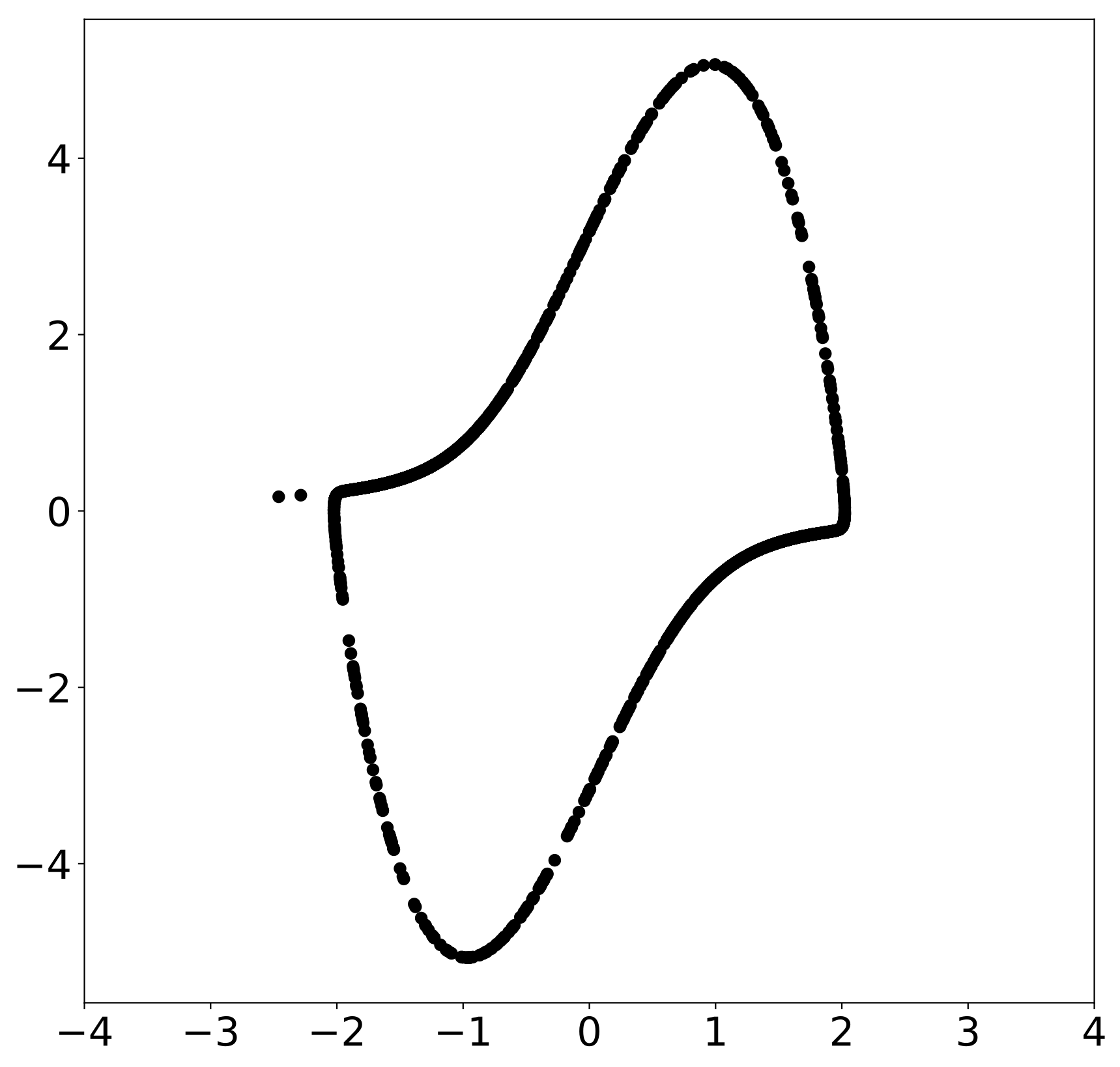}
\caption{ }\label{vdp_250_evol_1250_2500_true}
\end{subfigure}
\caption{Predictions of 2000 trajectories from the holdout set at time $T$ for Van der Pol system: (a)~and (b) are produced by RN1 for starting times $T/2$ and $T/4$ respectively, (c)~is reference solution produced by the ODE solver} 
\label{fig:vdp_pred}
\end{figure}

\begin{figure}
\centering
\begin{subfigure}[b]{.49\linewidth}
\includegraphics[width=\linewidth]{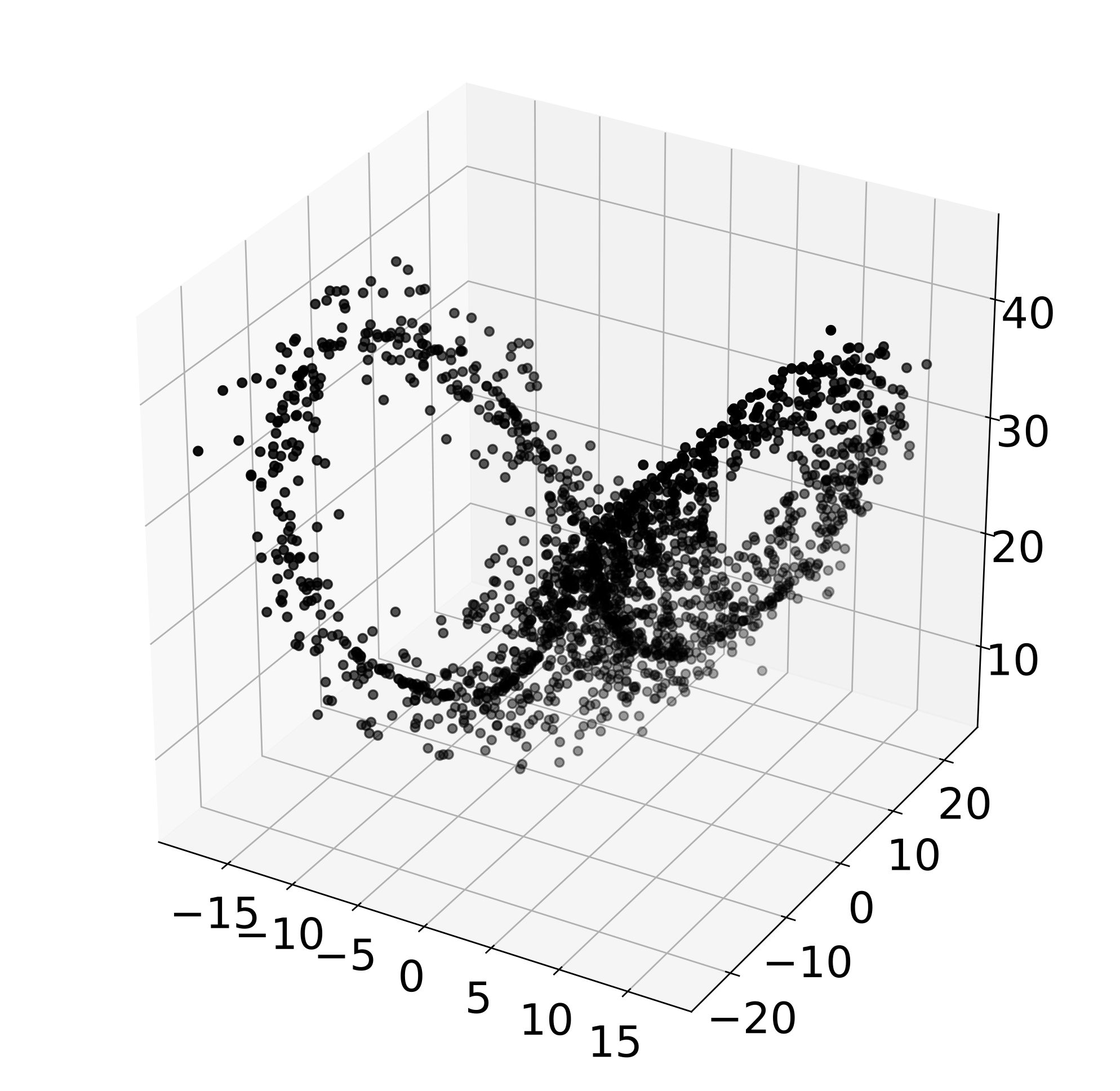}
\caption{ }\label{lorenz2_250_evol_1250_2500_pred}
\end{subfigure}
\begin{subfigure}[b]{.49\linewidth}
\includegraphics[width=\linewidth]{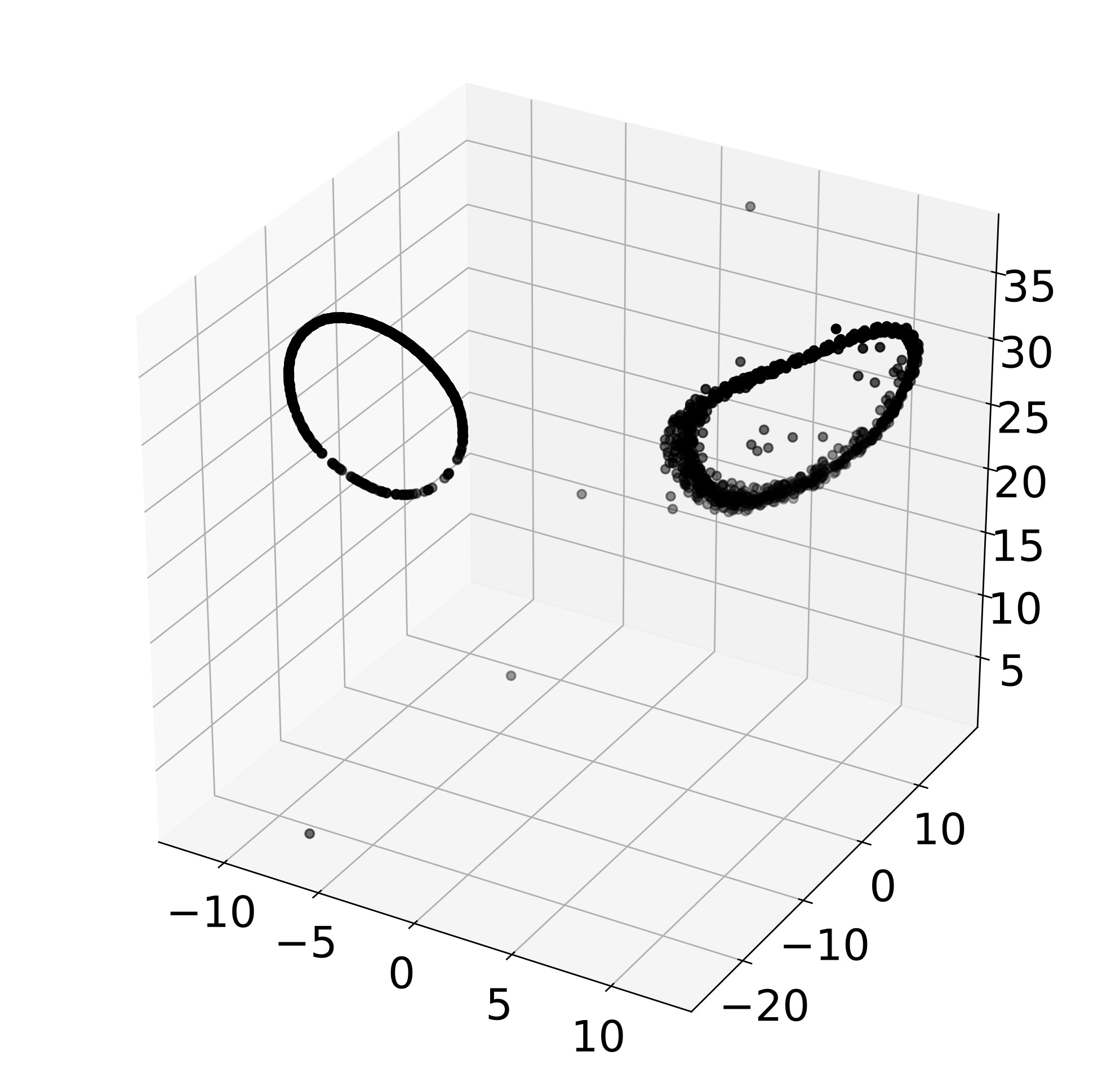}
\caption{ }\label{lorenz2_125_evol_625_2500_pred}
\end{subfigure}
\begin{subfigure}[b]{.49\linewidth}
\includegraphics[width=\linewidth]{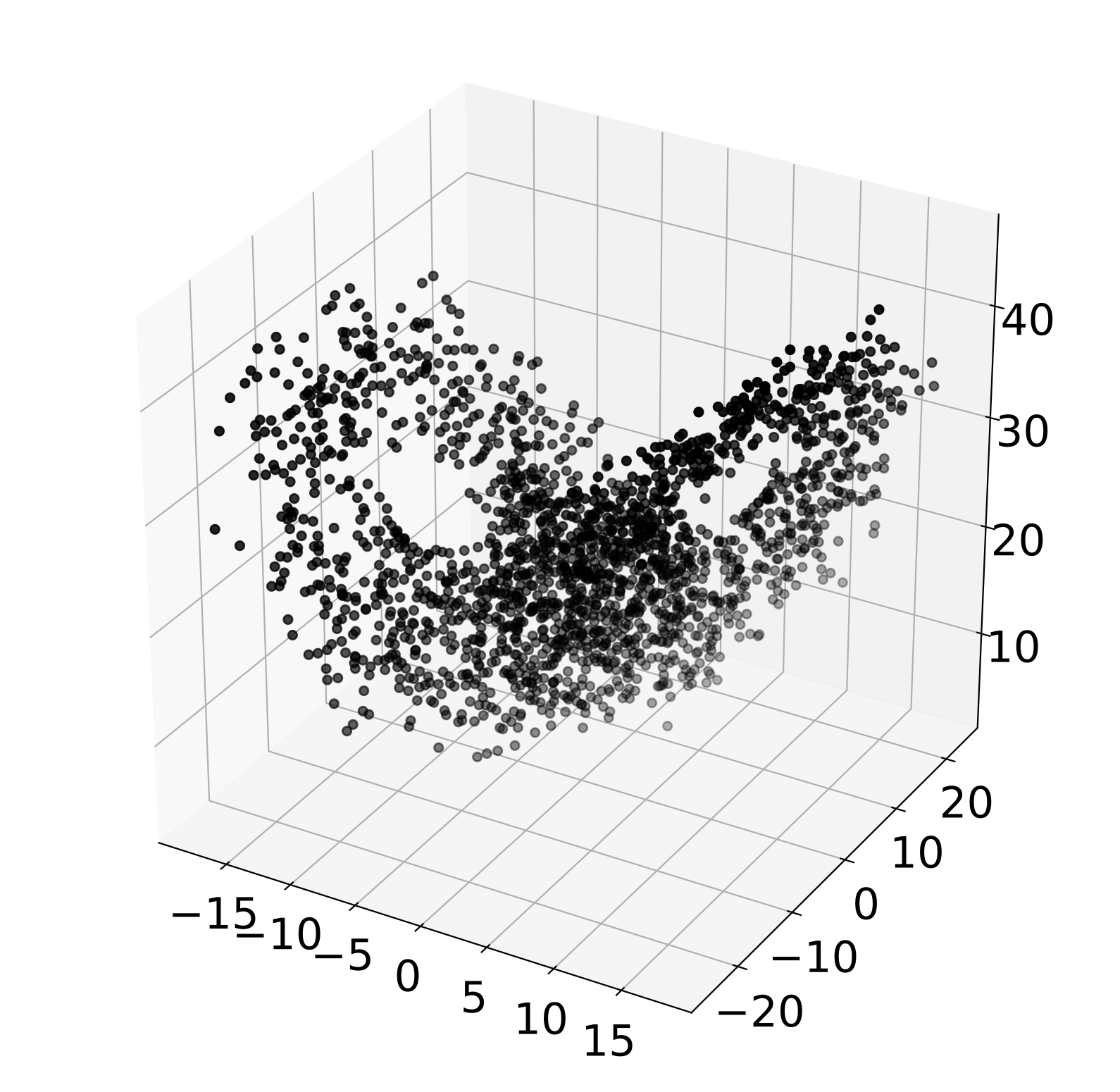}
\caption{ }\label{lorenz2_250_evol_1250_2500_true}
\end{subfigure}
\caption{Predictions of 2000 trajectories from the holdout set at time $T$ for Lorenz system: (a)~and (b) are produced by RN2 for starting times $T/2$ and $T/4$ respectively, (c)~is reference solution produced by the ODE solver} 
\label{fig:lorenz_pred}
\end{figure}

\begin{figure}
\centering
\begin{subfigure}[b]{.49\linewidth}
\includegraphics[width=\linewidth]{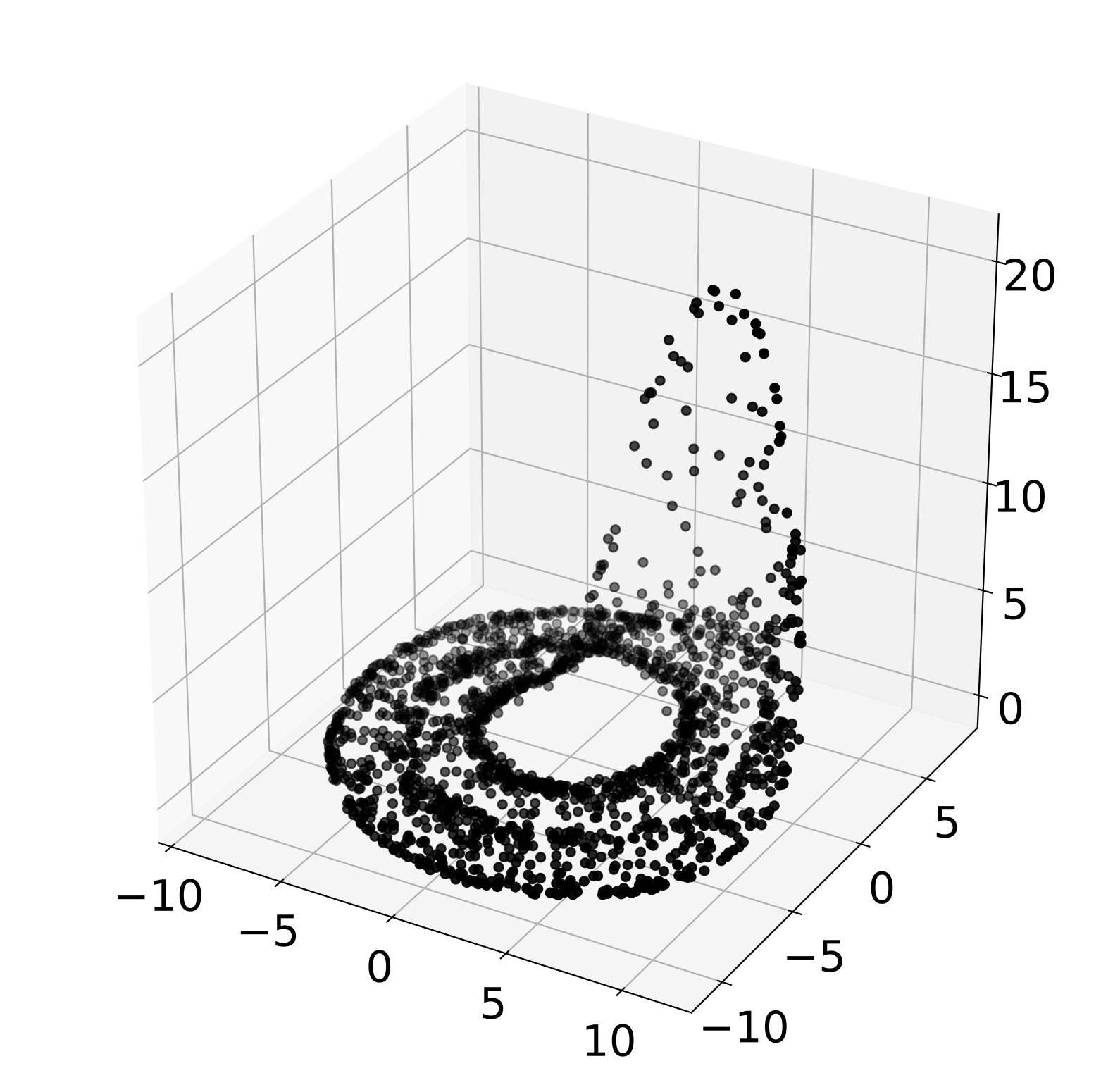}
\caption{ }\label{rossler_250_evol_1250_2500_pred}
\end{subfigure}
\begin{subfigure}[b]{.49\linewidth}
\includegraphics[width=\linewidth]{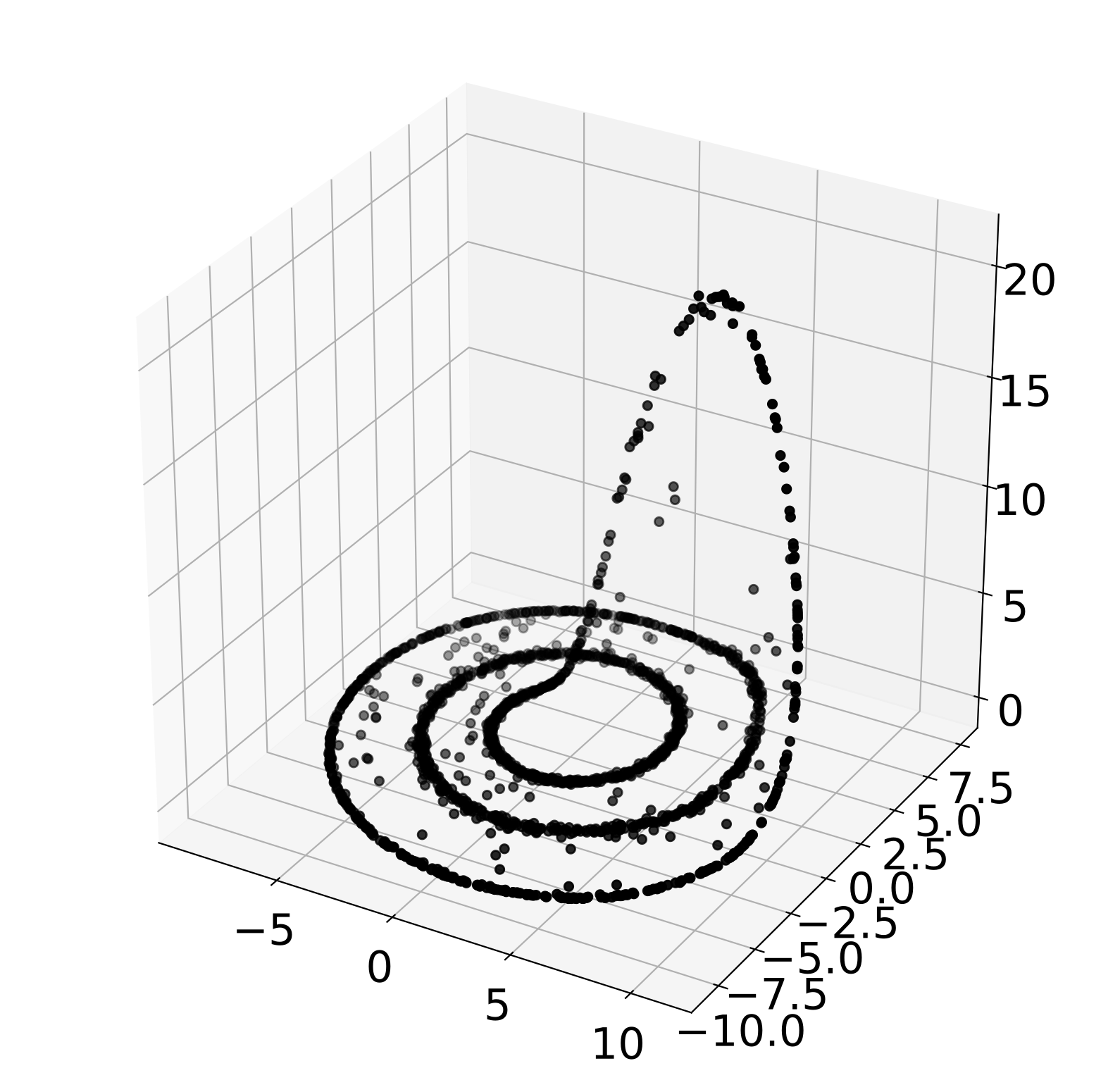}
\caption{ }\label{rossler_125_evol_625_2500_pred}
\end{subfigure}
\begin{subfigure}[b]{.49\linewidth}
\includegraphics[width=\linewidth]{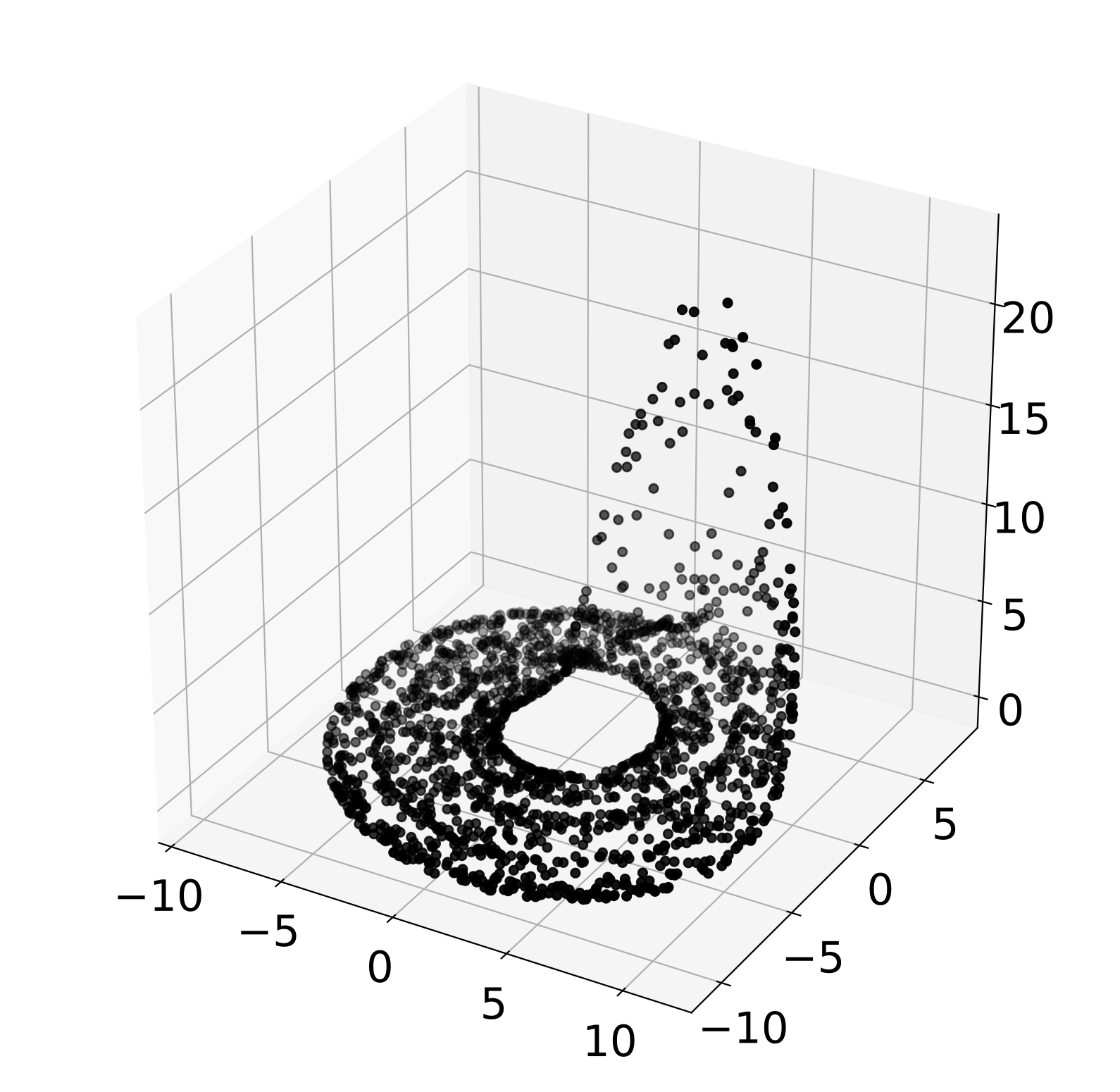}
\caption{ }\label{rossler_250_evol_1250_2500_true}
\end{subfigure}
\caption{Predictions of 2000 trajectories from the holdout set at time $T$ for R\"ossler system: (a)~is produced by RN3 from the time $T/2$, (b) is the prediction of RN1 for starting time $T/4$,
(c)~is reference solution produced by the ODE solver} 
\label{fig:rossler_pred}
\end{figure}

\begin{figure}
\centering
\begin{subfigure}[b]{.49\linewidth}
\includegraphics[width=\linewidth]{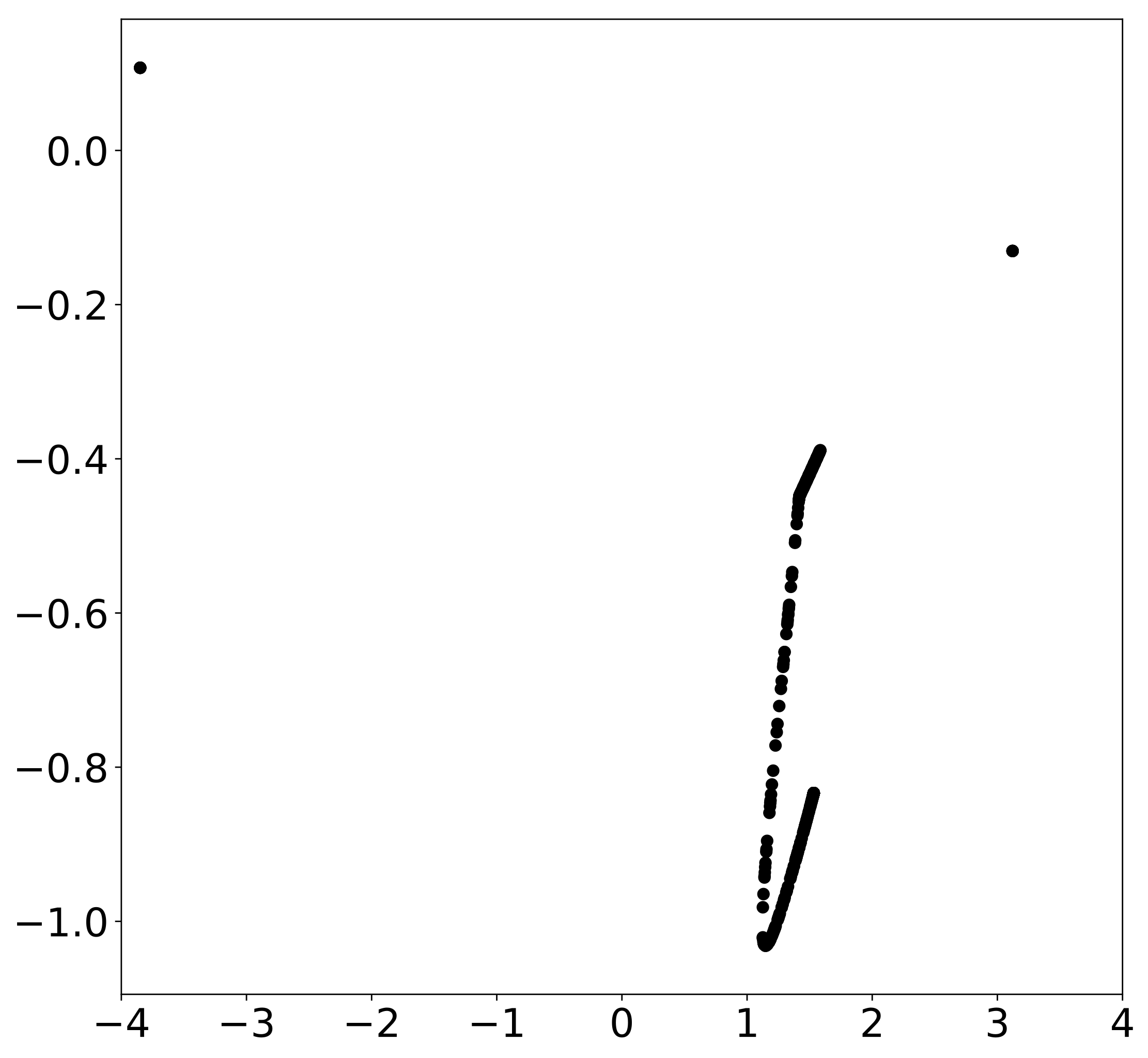}
\caption{ }\label{vdp_nonrn_250_evol_1250_2500_pred}
\end{subfigure}
\begin{subfigure}[b]{.49\linewidth}
\includegraphics[width=\linewidth]{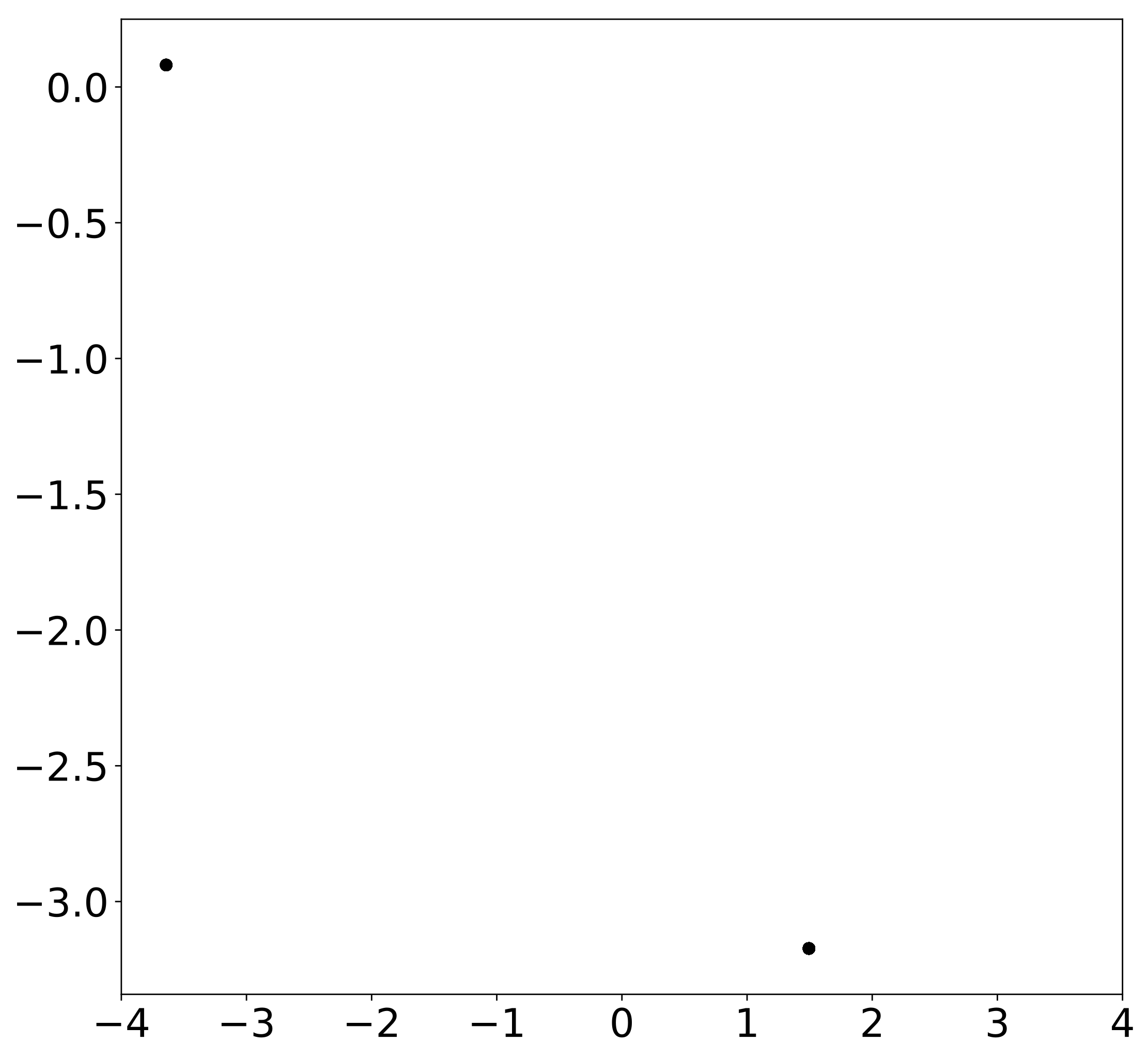}
\caption{ }\label{vdp_nonrn_125_evol_625_2500_pred}
\end{subfigure}
\begin{subfigure}[b]{.49\linewidth}
\includegraphics[width=\linewidth]{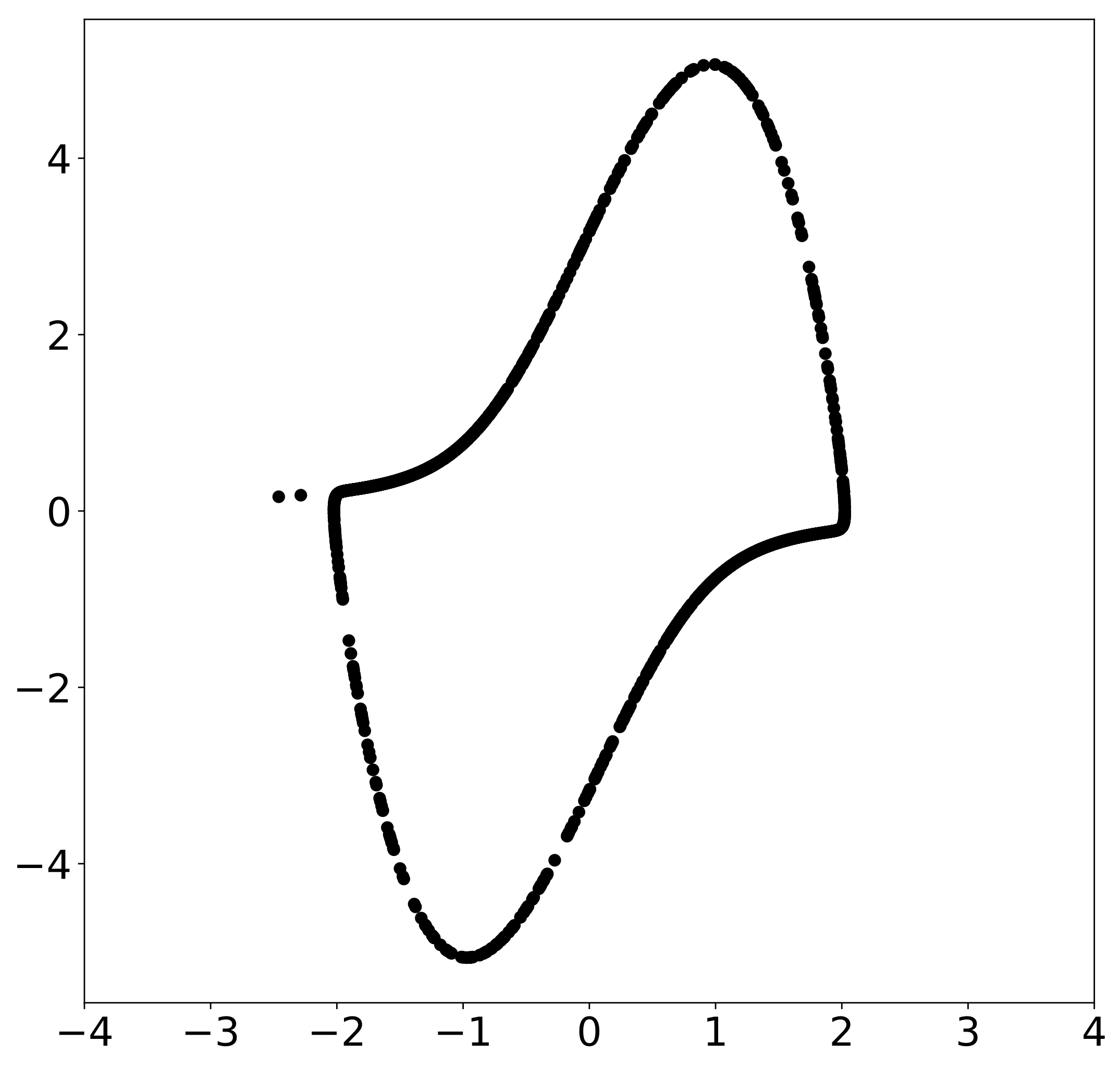}
\caption{ }\label{vdp_nonrn_250_evol_1250_2500_true}
\end{subfigure}
\caption{Predictions of 2000 trajectories from the holdout set at time $T$ for Van der Pol system: (a)~and (b) are produced by MLP for starting times $T/2$ and $T/4$ respectively, (c)~is reference solution produced by the ODE solver} 
\label{fig:vdp_mlp_pred}
\end{figure}

For the Van der Pol oscillator ODE system, all the four networks exhibit favorable results (see Figure~\ref{fig:vdp_errors}). They reproduce the evolution of the system with a good accuracy.
As can be seen in Table~\ref{tab:errs1}, the
smallest prediction errors for the interval $[T/2,T]$ are attained
by RN1, with $\varepsilon_{\text{avg}} \approx 0.61$, and
by RN2, with $\varepsilon_T \approx 1.07$.
For $[T/4,T]$, the best result for both errors is given by RN1,
with $\varepsilon_{\text{avg}} \approx 0.42$ and $\varepsilon_T \approx 0.73$ (see Table~\ref{tab:errs2}).
To visualize the network reconstruction quality, in
Figure~\ref{fig:vdp_pred} we compare the predictions of the ODE solver and
ResNet for $2\,000$ holdout snapshots at time $T$.
As we see in the figure, the networks have successfully reproduced the main curve, with just a small number of outlier points on the left.

The Lorenz system turns out to be the most difficult for the networks to learn. One can see in Figure~\ref{fig:lorenz_errors} and Tables~\ref{tab:errs1}-\ref{tab:errs2} that some experiments result in
large
prediction errors and even overflow (shown in the table as ``NaN'').
One of the possible reasons for this could be the chaotic nature of the system, which prevents the networks from accurately learning it. The error accumulates over time, and the predicted trajectories significantly diverge from those
produced by the ODE solver.
Nonetheless, some network architectures manage to show a reasonable performance. For predictions in the interval $[T/2,T]$, the network with the smallest errors is RN2, with $\varepsilon_{\text{avg}} \approx 0.38$ and $\varepsilon_T \approx 0.57$.
RN2 also turns out to be the best for the interval $[T/4,T]$,
with $\varepsilon_{\text{avg}} \approx 0.34$ and $\varepsilon_T \approx 0.58$. Comparison of predictions by solver and networks in Figure~\ref{fig:lorenz_pred} showcases the problems mentioned above. The RN2 predictions from $T/4$ seem to capture the revolution of points around the attractor properly, but they
are unable to reproduce the main butterfly-shaped dynamics of the system.
However, in case of predicting from $T/2$ RN2 shows far better results that closely resemble the solver predictions.

Finally, we evaluate the networks on the R\"ossler attractor.
All of them show considerably stable results, which is expected
because the evolution of the system is simpler to learn than in
case of Lorenz system.
RN3 yields the smallest errors for $[T/2,T]$,
with $\varepsilon_{\text{avg}} \approx 0.31$ and $\varepsilon_{T} \approx 0.53$,
whereas RN1 shows the best performance for $[T/4, T]$,
with $\varepsilon_{\text{avg}} \approx 0.42$ and $\varepsilon_{T} \approx 0.65$. Visual analysis of predictions at time~$T$ (see Figure~\ref{fig:rossler_pred}) shows that the networks in both experiments successfully reconstruct the form of the attractor, but in case of prediction from $T/4$ the points
exhibit a wrong tendency
to concentrate
near the horizontal spiral.

We now compare the performance of ResNet to MLP on the problem of predicting Van~der~Pol oscillator trajectories. It can be seen from Figure~\ref{fig:vdp_errors} that the relative error values are comparable to those of RN1--RN4. The numerical values are $\varepsilon_{\text{avg}} \approx 1.03$, $\varepsilon_{\text{T}} \approx 1.13$ for prediction from $T/2$ and $\varepsilon_{\text{avg}} \approx 1.81$, $\varepsilon_{\text{T}} \approx 2.15$ for prediction from $T/4$. The main difference is that in MLP's case the error quickly rises to a certain value and keeps oscillating around it further in time. However, the prediction plots clearly showcase inability of MLP to predict the evolution for long time intervals.
Except for a small amount of points that still evolve by the time $T$, most of them have gathered in several clusters and remain almost motionless (see Figure~\ref{fig:vdp_mlp_pred}).
That lack of motion in the trajectories explains the error stagnation seen on the plot for MLP.

\begin{table}
\centering
  \caption{Prediction errors for the interval $[T/2, T]$. The smallest
  error in each column is printed in boldface.}
  \label{tab:errs1}
  \begin{tabular}{| l | l | l | l | l | l | l |}
    \hline
        & \multicolumn{2}{c|}{Van der Pol} & \multicolumn{2}{c|}{Lorenz} & \multicolumn{2}{c|}{R\"ossler} \\ \cline{2-7}
        & $\varepsilon_{\text{avg}}$ & $\varepsilon_{T}$ & $\varepsilon_{\text{avg}}$ & $\varepsilon_{T}$ & $\varepsilon_{\text{avg}}$ & $\varepsilon_{T}$ \\ \hline

    RN1 & \textbf{0.61} & 1.09 & $1.60\text{e}17$ & $6.80\text{e}18$ & 0.39 & 0.62 \\ \hline

    RN2 & 0.83 & \textbf{1.07} & \textbf{0.38} & \textbf{0.57} & 0.97 & 1.69 \\ \hline

    RN3 & 0.70 & 1.27 & 0.52 & 1.19 & \textbf{0.31} & \textbf{0.53} \\ \hline

    RN4 & 1.01 & 1.49 & 0.42 & 0.57 & 0.32 & 0.59 \\
    \hline
  \end{tabular}
\end{table}

\begin{table}
\centering
  \caption{Prediction errors for the interval $[T/4,T]$. The smallest
  error in each column is printed in boldface.}
  \label{tab:errs2}
  \begin{tabular}{| l | l | l | l | l | l | l |}
    \hline
        & \multicolumn{2}{c|}{Van der Pol} & \multicolumn{2}{c|}{Lorenz} & \multicolumn{2}{c|}{R\"ossler} \\ \cline{2-7}
        & $\varepsilon_{\text{avg}}$ & $\varepsilon_{T}$ & $\varepsilon_{\text{avg}}$ & $\varepsilon_{T}$ & $\varepsilon_{\text{avg}}$ & $\varepsilon_{T}$ \\ \hline

    RN1 & \textbf{0.42} & \textbf{0.73} & nan & nan & \textbf{0.42} & \textbf{0.65} \\ \hline

    RN2 & 7.84 & 90.51 & \textbf{0.34} & \textbf{0.58} & 0.65 & 0.72 \\ \hline

    RN3 & 0.63 & 1.16 & 0.41 & 0.59 & 1.33 & 1.82 \\ \hline

    RN4 & 1.47 & 1.66 & nan & nan & 0.49 & 0.88 \\
    \hline
  \end{tabular}
\end{table}

\section{Conclusions and Future work}
\label{sec:future}

In this paper we show that neural networks can be
successfully used
to predict solution of ODE systems $\dot{\bm{x}}(t)=F(\bm{x}(t))$
based on solution snapshots only, i.e., without knowing the right
hand side function $F(\bm{x})$.
The key point of our approach is that the residual neural networks
(ResNet)
are employed, which are designed to avoid vanishing and exploding gradients
in deep networks and appear to be very suitable for our purpose.
Evaluation of ResNet on three test nonlinear ODE systems with
a chaotic behavior demonstrate their ability to capture the dynamics of the systems sufficiently well. The tests also show
excellent stability properties of ResNet, which allows prediction for
much longer time intervals than is currently possible with
other comparable machine learning approaches.

Several directions of further research can be indicated.
First, ResNet should be tested on more difficult realistic practical problems. Second, various modifications of residual networks, e.g., RevNet,
have recently been designed~\cite{gomez2017reversible}
and should be considered for evaluation.
As certain properties of such networks indicate,
an improved prediction accuracy can be reached with RevNet and alike.



\bibliography{elsarticle-main}

\end{document}